\documentclass[11pt,oneside,letterpaper]{article}
\usepackage{amssymb}
\usepackage{amsmath}
\usepackage[dvips]{graphicx}
\usepackage{setspace}
\usepackage{fancyhdr}
\usepackage{xcolor}
\usepackage{ifpdf}
\usepackage{graphicx}
\usepackage{rotating}
\usepackage{comment}
\usepackage{braket}
\usepackage{bbold}
\usepackage[pdfusetitle,bookmarks,pdfpagelabels,breaklinks,plainpages=false,pdfpagemode=UseNone]{hyperref}
\usepackage{cleveref}
 \usepackage[utf8x]{inputenc}

\usepackage{graphicx} % Required for inserting images
\usepackage{amssymb}
\usepackage[a4paper, left=3cm, right=3cm, top=3cm, bottom=3cm]{geometry}
\usepackage{amsmath}
\usepackage{tikz}
\usepackage{index}
\usepackage{verbatim}
\usepackage{amssymb}
\usepackage{physics}
\usepackage{blindtext}
\usepackage{graphicx}
\usepackage{caption}
\usepackage{subcaption}
\usetikzlibrary{calc,arrows.meta}
\usepackage{slashed}
\usepackage{mathtools}
\usepackage{mathrsfs}
\usepackage{hyperref}

\usepackage{datetime,color}
\usepackage{graphicx}
\usepackage{float}
\usepackage{authblk}
\usepackage[title]{appendix}

\newcommand{\be}{\begin{equation}}

\newcommand{\ee}{\end{equation}}
\newcommand{\nin}{\noindent}

\def\IC{\relax\,\hbox{$\inbar\kern-.3em{\rm C}$}}

\def\cA{{\cal A}}

\def\cF{{\cal F}}

\def\cL{{\cal L}} 
 \def\cO{{\cal O}}
 
\def\cR{{\cal R}} 
 
\def\nin{\noindent}
\def\a{\alpha}
\def\b{\beta}

\def\vf{\varphi}
\def\F{\Phi}
\def\c{\chi}
\def\g{\gamma}
\def\G{\Gamma}
\def\s{\sigma}
\def\e{\epsilon}
\def\f{\phi}
\def\ve{\varepsilon}
\def\l{\lambda}
\def\z{\zeta}

\def\ua{\underline{a}}

\let\h\hat
\def\l{\left( }
\def\r{\right)}
\def\de{\partial}
\def\fr{\frac}
\def\d{\delta}
\def\D3{D3}

\def\tf{\tilde f}

\def\pfa{p_{\f_a}}
\def\ppa{p_{\psi_a}}

\title{\texorpdfstring{\vspace{20pt}}{}\textsc{On Grey Galaxies and D3-branes on the Conifold
} \texorpdfstring{\vspace{30pt}}{}}
\author[$\sharp$ $\dag$]{Antonio Amariti}
\author[$\sharp$ $\dag$]{Riccardo Salto}
\author[$\Diamond$ $\heartsuit$]{Chiara Toldo \vspace{10mm}}

\affil[$\sharp$]{\it \footnotesize INFN, Sezione di Milano, Via Celoria 16, I-20133 Milano MI, Italy \vspace{5mm}}

\affil[$\dag$]{\it \footnotesize Dipartimento di Fisica, Universit\`a di Milano, via Celoria 16, I-20133 Milano MI, Italy \vspace{5mm}}

\affil[$\Diamond$]{\it \footnotesize Physique Th\'eorique et Math\'ematique and International Solvay Institutes, Universit\'e Libre de Bruxelles, C.P. 231, 1050 Brussels, Belgium \vspace{5mm}}

\affil[$\heartsuit$]{\it \footnotesize Departament de Matem\`atiques \& SYMCREA, Universitat Polit\`ecnica de Catalunya, Av. Dr. Mara\~{n}\'on, 02028 Barcelona, Spain}

\date{}

\begin{document}
  \maketitle
  \vspace{100pt}

    \begin{abstract}

  \vspace{5mm}
  
    \noindent
We study the effective potentials for various probe D-branes surrounding AdS$_5$ black holes in consistent
truncations of type IIB supergravity on Sasaki–Einstein 5-manifolds $Y^{p,q}$, focusing in particular on the conifold base $T^{1,1}$. The probes are either
dual Giant Gravitons having support on AdS spacetime dimensions or D3 probe branes wrapped on
noncontractible cycles of the internal manifold, which then appear ”point-like” in the AdS spacetime. The non-contractible cycles of $T^{1,1}$ and $Y^{p,q}$ topologically stabilize the branes at a finite size, and we find stable supersymmetric wrapped D3 branes on Gutowski-Reall black holes, effectively providing a stable BPS composite system in AdS$_5$ made by a black hole surrounded by a baryonic condensate. We show that the onset of superradiance in the black hole backgrounds matches the emergence of stable minima in the probe effective potentials, extending the constructions of Dual Dressed Black Holes and the Grey Galaxies. We comment on the relevance of
this result for recent studies concerning the classical cohomology problem and the associated fortuitous states in the dual Klebanov-Witten theory. 

\end{abstract}

  \newpage
  \setcounter{tocdepth}{2}
  \tableofcontents
      
   \section{Introduction}\label{Intro}

   Within the AdS/CFT framework, asymptotically Anti-de Sitter (AdS) black holes play a central role, providing holographic descriptions of thermal states and finite-temperature phenomena in the dual quantum field theory. Their thermodynamic and dynamical properties therefore offer a valuable window into the phase structure of strongly coupled systems. 
   
   Importantly, however, black hole solutions are not generically stable throughout their parameter space. Much like ordinary thermodynamic systems, they may develop instabilities, which can signal the emergence of new phases. From the gravitational perspective, such instabilities are often associated with development of a nontrivial matter profile outside the horizon, giving rise to black hole solutions with scalar “hair”. In the dual field theory, these processes are interpreted as the onset of phenomena such as spontaneous symmetry breaking and superconducting or superfluid phases. Understanding the mechanisms that trigger these instabilities, as well as the structure of the resulting phases, is therefore an important step toward elucidating the rich phase structure and dynamics of holographic quantum field theories.

    It is a well-known fact that Kerr black holes in $AdS_d$ spacetime, with $d \geq4$ present classical superradiant instabilities \cite{Cardoso:2004hs}. In a nutshell, superradiance is the process by which a rotating black hole can amplify certain waves and, if those waves are trapped, develop an instability. More concretely, if a wave
    \begin{equation}
        \Phi \sim e^{i \omega \tau + i m \phi}
    \end{equation}
    scatters on a rotating black hole with horizon angular velocity $\Omega$, if
    \begin{equation}
    \omega < m \Omega
    \end{equation}
    the wave scattered from the black hole can come out with more energy than it had initially, the black hole effectively losing angular momentum and energy.  If the wave is trapped (i.e. due to the presence of a potential, such as the Anti-de Sitter one), its amplitude grows unboundedly and this creates an instability.

    Recently, the authors of \cite{Kim:2023sig} conjectured that the endpoint of the superradiant instability is a so-called "Grey Galaxy" (GG): a central black hole surrounded by a macroscopic “galactic” cloud/disk of rotating gas. This is motivated by the fact that the AdS potential confines the  angular momentum extracted from the black hole via superradiance, and then one needs an equilibrium configuration in which the excess angular momentum is stored outside the black hole. In the Grey Galaxy configuration, the gas can carry a finite fraction of the total energy and angular momentum. At large separation, the interactions between the black hole and the gas are parametrically suppressed, allowing the two components to be treated approximately independently.
     The equilibrium configuration is selected by maximizing the total entropy subject to fixed total energy and angular momentum.

    A similar mechanism happens for charged superradiance: in this case there is a charged analogue of the Grey Galaxy story, where the black hole charge is stored outside the black hole by means of dual Giant Gravitons\footnote{dual Giant Gravitons are particular extended D-brane configurations that wrap around a contractible cycle inside an Anti-de Sitter (AdS) spacetime. In comparison, Giant Gravitons \cite{Grisaru:2000zn} are D-brane configurations that wrap cycles in the internal manifold, and are usually stabilized by angular momentum against their tendency to shrink.}, hence forming a so-called "Dual Dressed black hole" (DDBH) \cite{DDBH}. Charged, rotating black holes in AdS can preserve supersymmetry: it is indeed the connection to supersymmetric black holes that turns the Grey Galaxy/DDBH picture from a classical endpoint-of-instability story into a microscopic counting problem. A DDBH can preserve supersymmetry, and taking advantage of the large separation between the two components, one can approximately factorize their contributions to the state counting and look for signatures of these configurations in the dual superconformal index \cite{Choi:2025lck,Deddo:2025jrg}.

    In this paper, we build on the Grey Galaxy and Dual Dressed Black Holes story, and investigate charge and angular momentum superradiance in type IIB compactifications on $AdS_5\times T^{1,1}$ and, more generally, on $AdS_5\times Y^{p,q}$, while revisiting the maximally supersymmetric $AdS_5\times S^5$ case from a different perspective. Crucially, compactifications on $T^{1,1}$, and more generally on $Y^{p,q}$, are characterized by non-contractible cycles, which stabilize Giant Gravitons at finite size, leading to a gas of extended D3-branes. In contrast, in the $S^5$ case no such topological stabilization is available, and the Giant Graviton minimizes its energy by collapsing to a pointlike graviton. From this perspective, our interpretation is that the nature of the emitted gas depends crucially on the topology of the internal manifold.

    Specifically, we consider type IIB supergravity on backgrounds of the form $AdS_5 \times X^5$, where $X^5$ is a $Y^{p,q}$ manifold, focusing on the $T^{1,1}$, the base of the conifold.  The latter has the topology of $S^2 \times S^3$ and can be described as a non-trivial $U(1)$ fibration over $S^2 \times S^2$. Unlike the sphere, $T^{1,1}$ possesses non-trivial three-cycles and the resulting gravitational theory is dual to the Klebanov-Witten $\mathcal{N}=1$ superconformal $SU(N) \times SU(N)$ gauge theory \cite{Klebanov:1998hh}. Concretely, we analyze the dynamics of dual Giant Gravitons and D3 branes wrapped on noncontractible cycles in the internal dimensions in ten-dimensional uplifts of five-dimensional black hole solutions. Only a few examples of black hole solutions of the corresponding supergravity truncation are known, for instance the static black brane solutions of \cite{Herzog:2009gd,Henriksson:2019ifu} and the supersymmetric near horizon geometries of \cite{Benini:2020gjh}. For our purposes, the full black hole geometry is needed, along with the presence of nonvanishing angular momentum. Therefore, our background solutions will be the rotating AdS$_5$ black holes \cite{Chong:2005hr} that have a smooth BPS limit \cite{Gutowski:2004ez}. These solutions live in the so-called universal truncation, which corresponds to minimal five-dimensional gauged supergravity.

    We show that the onset of charge and angular momentum superradiance is associated with the appearance of stable minima in the corresponding probe effective potentials, thereby extending the probe-brane picture of \cite{DDBH} beyond the Dual Dressed Black Hole construction, and providing a probe-brane realization of the Grey Galaxy proposal of \cite{Kim:2023sig,Choi:2025lck}. At the onset of both charge and angular momentum superradiance, we find stable minima of the probe effective potentials located at large distances from the black hole horizon. In the supersymmetric limit, these minima become marginally stable and satisfy the corresponding BPS bounds, indicating the existence of supersymmetric bound states between the probe and the central black hole. We verify the supersymmetry of these configurations through the $\kappa$-symmetry analysis of the probe brane embeddings. This effectively provides a stable BPS composite system in AdS spacetime, whose construction is notoriously difficult\footnote{Evidence for the existence of multicenter  black holes in Anti-de Sitter spacetimes in the probe approximation (extreme mass ratio) was given in \cite{Anninos:2013mfa,Monten:2016tpu,Monten:2021som} for AdS$_4$ spaces, where certain "glassy" properties \cite{Anninos:2011vn,Anninos:2012gk} of these systems were also analyzed. The possible existence of \textit{supersymmetric} bound states in AdS made by wrapping D-branes however was not established and is a genuinely new feature of the present work.}, made by a black hole and a stable wrapped D3 brane probe ("baryonic condensate"). An example of such bound state is depicted in Fig. \ref{fig:GGT11SUSYclpot}, which can be regarded as the main result of the paper.

While completing this draft, \cite{Choi:2026faq} anticipated part of our results from the dual field theory side via the analysis of fortuitous cohomologies in the conifold. In particular, a prediction was given for the presence of black hole solutions surrounded by a baryonic condensate. In their setup this condensate arises from a macroscopic baryonic Higgs VEV that resolves the conifold singularity, interpolating between an  asymptotic AdS$_5\times T^{1,1}$ region in the UV and the AdS$_5\times S^5$ throat in the IR.  At large VEV the two regions in the IR are weakly coupled, suggesting a hairy black hole description of  the kind studied in our paper,  while for small VEV it is expected that they merge into a  BPS black hole in AdS$_5\times T^{1,1}$. We will elaborate on this more in the conclusions, and as we will see our results are compatible with their prediction, but we believe a possible interpretation in terms of multiply centered black holes in the extreme mass ratio (AdS$_5$ black hole bound states) is possible as well.

    The paper is organized as follows. In section~\ref{sugra}, we review the five-dimensional supergravity model obtained from the reduction of type IIB supergravity on the Sasaki-Einstein manifolds $T^{1,1}$ and $S^5$, and present the relevant uplift formulae. We also discuss the corresponding black hole solutions and their supersymmetric limits. In section~\ref{spacetime_filling}, we generalize the Dual Dressed Black Hole construction as an endpoint of charge superradiance in $AdS_5\times T^{1,1}$. Section~\ref{GGT11} is devoted to the study of the effective potentials derived for D3-branes wrapped on non-contractible cycles of $T^{1,1}$ and their role as candidate endpoints of angular momentum superradiance. These considerations are generalized to $Y^{p,q}$ truncations in section ~\ref{WrappedYpq}. In section~\ref{GGS5}, we revisit the maximally symmetric $AdS_5\times S^5$ case and derive the Grey Galaxy construction from the probe-brane perspective. The computations relevant for the $\kappa$-symmetry analysis are in Appendix (\ref{ksymmetry}).  We conclude by summarizing the main results, connection with recent literature, and directions to explore in the near future.

\section{Five-dimensional black holes and their 10d uplift \label{sugra}}
    We present the relevant AdS$_5$ black hole solution that will be the background probed by the D3-branes. These are general rotating black holes solutions with temperature \cite{Cvetic:2004hs,Chong:2005hr} of minimal 5d $U(1)$ gauged supergravity which admit a regular supersymemtric limit \cite{Gutowski:2004ez}. We restrict our treatment to the case of equal angular momenta, that exhibit symmetry enhancement. We present here their uplift to 10d as well.
    
\subsection{Black hole solutions of five-dimensional minimal gauged supergavity}
    We consider five-dimensional minimal gauged supergravity, whose bosonic sector consists in the metric and the graviphoton. The Lagrangian is
    \begin{equation}
        \label{lagMGS}
        e^{-1}\cL= \cR+\fr{12}{L^2}-\fr34F_{\mu\nu}F^{\mu\nu}+\fr14\ve^{\mu\nu\rho\s\lambda}F_{\mu\nu}F_{\rho\s}A_\lambda\,,
    \end{equation}
    where $\cR$ is the 5d Ricci scalar.
    
    We work with the charged, rotating, black hole solution first presented in \cite{Cvetic:2004hs} and subsequently analyzed in \cite{Madden:2004ym}:
    \begin{equation}
        \begin{split}\label{5dsolution}
          ds_5^2&= -\frac{r^2 W(r)}{4b(r)^2}dt^2+\frac{dr^2}{W(r)} + \frac{r^2}{4}(\sigma_1^2+\sigma_2^2)+b(r)^2(\sigma_3+f(r)\, dt)^2,\\
                A &= \frac{q}{r^2}\left(dt -\frac{j}{2}\sigma_3 \right),
        \end{split}     
    \end{equation}
    where \begin{equation}\label{sigmas}
        \begin{split}
                \sigma_1 &= \cos{\psi_a} d\theta_a +\sin{\psi_a}\sin{\theta_a} d\phi_a,\\
                \sigma_2 &= -\sin{\psi_a} d\theta_a + \cos{\psi_a}\sin{\theta_a} d\phi_a,\\
                \sigma_3 &=d\psi_a + \cos{\theta_a} d\phi_a,
        \end{split} 
    \end{equation}
    are $SU(2)$ right-invariant one-forms. The functions in the metric are defined
    \begin{equation}\label{5dsolfunctions}
        \begin{split}
                b(r)^2 & =\frac{r^2}{4}\left(1-\frac{j^2 q^2}{r^6}+\frac{2 j^2 p}{r^4}\right) ,\quad
                f(r)  =-\frac{j}{2 b^2}\left(\frac{2 p-q}{r^2}-\frac{q^2}{r^4}\right),\\
                W(r) & =1+\fr4{L^2} b^2-\frac{1}{r^2}\left(2 p-2 q\right)+\frac{1}{r^4}\left(q^2+2pj^2\right).
        \end{split} 
    \end{equation}
    The coordinate range is $t\in \mathbb{R},\,\,r\in[0,\infty),\,\,\theta_a\in[0,\pi],\,\,\f_a\in[0,2\pi),\,\,\psi_a\in[0,4\pi).$ 
    The solution describes an asymptotically AdS$_5$ black hole with equal angular momenta on the spatial $S^3$ and one electric charge. It is characterized by three independent quantities parameterizing the mass, the electric charge, and the angular momentum:
    \begin{equation}\label{5dsolcharges}
       M=\frac12\left(3p-3q+\fr{pj^2}{L^2} \right),  \quad Q=\frac q2,\quad  J=\frac j2\left( 2p-q\right).
   \end{equation}
    The thermodynamic chemical potentials conjugate to angular momentum and electric charge are given by 
    \begin{equation}\label{5dsolchemicalpot}
        \Omega=f(R),\quad \mu=A_t(R)-\Omega A_{\sigma_3}(R),
    \end{equation}
    with $R$ the radius of the outer event horizon, i.e. the largest root of $W(r)=0$.
    The Hawking temperature is 
    \begin{equation}\label{5dsoltemperature}
        T=\frac{R\, W'(R)}{8\pi \,b(R)}. 
    \end{equation}
    In the gauge choice of \eqref{5dsolution}, the graviphoton field vanishes asymptotically but is not regular at the horizon in Eddington-Finkelstein-like coordinates. Regularity is restored by the shift
    \begin{equation}
        A_t\mapsto A_t-\mu.
    \end{equation}
    
\subsubsection{The supersymmetric limit}
    Supersymmetry and extremality impose a set of relations among the parameters of the general, non-extremal solution:
    \begin{equation}\label{5dGRsolsusycond}
            p=2 R^2 \left( 1+\frac{R^2}{2 L^2}\right)^2,\quad
            q=R^2 \left( 1+\frac{R^2}{2L^2}\right),\quad
            j=\frac{R^2}{2L} \left( 1+\frac{R^2}{2L^2}\right)^{-1},
    \end{equation}
    which lead to the Gutowski-Reall black hole \cite{Gutowski:2004ez}. The BPS bound in these conventions is 
    \begin{equation}
        M=3 Q+2 \frac{J}{L}.
    \end{equation}
    The metric reads
    \begin{equation}\label{5dGRsolMinwallacoords}
        \begin{split}
            ds_5^2  &=  - \frac{U}{\Sigma} dt^2 + \frac{ dr^2}{U} + \frac{r^2}{4} \left( \sigma_{1}^2 + \sigma_{2}^2 +\Sigma\,\left(\sigma_3-\Omega\,dt\right)^2 \right),\\
            A &=  \left(  \frac{R^2}{r^2} +\frac{R^4}{2 L^2\,r^2}\right)dt- \frac{R^4}{4\,L\, r^2}\sigma_3 ,
        \end{split}
    \end{equation}
     where
    \begin{equation}\label{5dGRfuncdef}
        \begin{split}
         U &= \left( 1 - \frac{R^2}{r^2} \right)^2 \left( 1 + 2 \frac{R^2}{L^2} + \frac{r^2}{L^2}\right), \quad
        \Sigma = 1 +\frac{R^6}{L^2 r^4} - \frac{R^8}{4 \, L^2 r^6},\\
        \Omega&= \frac2{\Sigma L}\left(\left( \frac32+ \frac{R^2}{L^2}\right)\frac{R^4}{r^4}-\left(\frac12+\frac{R^2}{4L^2} \right) \frac{R^6}{r^6}\right).
        \end{split}
    \end{equation}
    In these coordinates the solution is manifestly AdS$_5$ as $r \rightarrow \infty$ since $g_{tt} \rightarrow r^2/L^2$, and $\Sigma =1$ and $\Omega =0$. The boundary does not rotate. Observe that by sending $L \rightarrow \infty$ one recovers the asymptotically flat supersymmetric solution, also known as the BMPV black hole \cite{Breckenridge:1996is}. 
    \nin
    In what follows, we will also make use of another coordinate system. We perform the following shift on the angular coordinate    \begin{equation}
        \psi_a\mapsto\psi_a+\fr2Lt,
    \end{equation}
    together with the gauge transformation $A_t\mapsto A_t-\mu$. The metric and gauge field then take the form  
   \begin{equation} \begin{split}\label{5dsolutionshifted}
          ds_5^2&= -\frac{r^2 W(r)}{4b(r)^2}dt^2+\frac{dr^2}{W(r)} + \frac{r^2}{4}(\sigma_1^2+\sigma_2^2)+b(r)^2(\sigma_3+\l f(r)+\fr2L\r dt)^2,\\
                A &= \l \frac{q}{r^2}\l1-\fr jL\r-\mu\r dt -\frac{q\,j}{2\,r^2}\sigma_3.
        \end{split}     
    \end{equation}
    In the extremal supersymmetric limit, we obtain \begin{equation}\label{5dGRsolution}
        \begin{split}
            ds_5^2  &=  - \tf^2 dt^2 - 2 \tf^2\, \Psi\, dt \sigma_3 + U^{-1} dr^2 + \frac{r^2}{4} \left( \sigma_{1}^2 + \sigma_{2}^2 +\Sigma\,  \sigma_3^2 \right),\\
            A &=  -\tilde f \,dt- \frac{R^4}{4\, L r^2}\sigma_3 ,
        \end{split}
    \end{equation}
   where
   \begin{equation}
        \tf = 1 - \frac{R^2}{r^2}, \quad
        \Psi = -\frac{r^2}{2L} \left( 1 + \frac{2 R^2}{r^2} + \frac{3 R^4}{2 r^2 (r^2- R^2)}\right),
   \end{equation}
and the $SU(2)$ right invariant forms are defined as in \eqref{sigmas} just in terms of the new $\psi_a'$. 

\subsection{Uplifts to type IIB supergravity}
    We now uplift the black hole solution to type IIB supergravity. This uplifted geometry will provide the background in which we study probe D3-brane configurations wrapping cycles in both the AdS$_5$ and internal directions. While the main text focuses on the conifold geometry $T^{1,1}$ and the five-sphere $S^5$, the mechanisms discussed in this work can be extended to the $Y^{p,q}$ manifolds, with the appropriate caveats, as we argue in section \ref{WrappedYpq}.
  \subsubsection{5d supergravity from IIB string theory on AdS$_5\times T^{1,1}$\label{5dSuGra}}
    The five-dimensional solution \eqref{5dsolution} admits a consistent uplift to type IIB supergravity on AdS$_5\times T^{1,1}$.   Here $T^{1,1}$ is viewed as a non-trivial $U(1)$ fibration over $S^2 \times S^2$. We parameterize the fiber with the coordinate $\psi\in[0,4\pi)$ and the coordinates on the two spheres with $\theta_{1}\in[0,\pi], \phi_1\in[0,2\pi)$ and $\theta_2\in[0,\pi], \phi_2\in[0,2\pi)$ respectively. Following \cite{Herzog:2009gd}, the ten-dimensional metric takes the form
    \begin{equation}\label{DDBHT11metric}
        \begin{split}
                       ds_{10}^2 &= ds_5^2 + L^2
                            \left(\frac{1}{6} \sum_{i=1,2}(d \theta_i^2 + \sin^2 \theta_i d\phi_i^2) 
                    + \frac{1}{9} \left(d \psi +\sum_{i=1,2} \cos \theta_id\phi_i  + \fr3L A\right)^2 \right),\\
        \end{split}
    \end{equation}
    where $ds_5^2$ and $A$ are the metric and gauge field \eqref{5dsolution}.
    The self-dual five-form is 
   \begin{equation}
    \begin{split}
        \fr{F_{(5)}}{L^4}&=\l 1+\star_{10} \r G_{(5)},\\
        G_{(5)}&=-{2  \over 27} \omega_2 \wedge \omega_2 \wedge g_5- {1\over 18L} F \wedge d g_5 \wedge g_5^A+{1 \over 18L} A \wedge dg_5 \wedge dg_5  ,\\
        \star_{10}\, G_{(5)}&= -\fr4{L^5} \text{vol}_5+ {1\over 6L^2} \star_5  F  \wedge dg_5  , 
    \end{split}
    \end{equation}
    where
    \begin{equation}
        \begin{split}
           g_5 =& d \psi + \cos \theta_1 d\phi_1 + \cos \theta_2 d\phi_2,\quad g_5^A = g_5 +\fr3LA ,\\
               \omega_2= &{1\over 2} \left(\sin \theta_1 d\theta_1 \wedge d\phi_1- \sin \theta_2 d\theta_2 \wedge d\phi_2 \right).
        \end{split}
    \end{equation}
    The five-dimensional graviphoton field enters the internal geometry through the shifted one-form $g_5^A$.\\
    A convenient gauge choice for the four-form Ramond-Ramond (R-R) potential satisfying $dC_{(4)}=F_{(5)}$ is
        \begin{equation}\label{fourfo}
            \frac{C_{(4)}}{L^4} = C_V- \frac{2}{27 }\psi \, \omega_2 \wedge \omega_2 -\frac{1}{6L^2}\star_5 F\wedge g_5-\frac{1}{18L} A\wedge dg_5 \wedge g_5^A,
        \end{equation}
        where $C_V$ is defined
        \begin{equation}
            \label{Cv}
            C_V= \frac{r^4-R^4}{8L^5}dt\wedge \sigma_1\wedge \sigma_2\wedge \sigma_3.
        \end{equation}
         The AdS radius is related to the color number $N$ via
    \begin{equation}\label{T11AdSradius}
        L^4=4\pi g_sN\l\a'\r^2\fr{27}{16}.
    \end{equation}

\subsubsection{5d supergravity from IIB string theory on AdS$_5\times S^5$}\label{UPLIFTS5SUSY}
        The maximally symmetric consistent truncation of type IIB supergravity is AdS$_5\times S^5$. We uplift the solution \eqref{5dsolution} using the coordinates of Appendix C in \cite{Aharony:2021zkr}, which will be useful for the $\kappa$-symmetry analysis of Appendix \ref{ksymmetry}. The ten-dimensional metric is
\begin{equation}\label{GGS510dmetric}
        \begin{split}
            ds^2&= ds^2_5 + L^2\left(d\rho_s^2+ \frac{1}{16} \sin^2(2\rho_s)\l d\z_s-\cos\theta_s d\f_s\r^2+\right.\\
            &\,\,\,\,\,\,\,\,\left.+\fr14\sin^2(\rho_s)\l d\theta_s^2 +\sin^2\theta_s d\f_s^2\r+\fr19\l d\psi_s+\cA+\fr3LA\r^2\right), \\
        \end{split}
    \end{equation}
where $ds_5^2$ and $A$ are defined in \eqref{5dsolution} and 
\begin{equation}
    \begin{split}
        &\cA= \fr32\sin^2(\rho_s) \l d\z_s-\cos\theta_s d\f_s\r-d\z_s.\\
    \end{split}
    \end{equation}
The five-sphere is written as a $U(1)$ fibration over $\mathbb{CP^2}$, with coordinates $\rho_s\in[0,\fr\pi2],\,\,\theta_s\in[0,\pi],\,\,\vf_s\in[0,2\pi),\,\,\zeta_s\in[0,4\pi),\,\,\psi_s\in[0,6\pi)$.
The Ramond-Ramond four-form potential is 
    \begin{equation}
        C_{(4)}=-\fr4{L^5}\b_{(4)}+\fr19 \l\cA+d\z_s\r\wedge\l\l\fr1{27} d\cA-\fr1LF\r\wedge \l d\psi_s+\cA+\fr3LA\r+\fr1{L^2}\star_{5}\,F\r,
    \end{equation}
where $d\b_{(4)}=\text{vol}_5$.
The AdS radius is 
\begin{equation}
    L^4=4\pi g_s N \l\a'\r^2.
\end{equation}
\nin
In the following, we set $L=1$.

\section{Black hole instabilities}

In \cite{DDBH,Kim:2023sig,Choi:2025lck}, the authors proposed that the endpoints of charge and angular momentum superradiance in AdS black holes are quasi-non-interacting mixed states consisting of a central black hole surrounded by either
    \begin{itemize}
        \item a small number of charged dual Giant Gravitons, the Dual Dressed Black Hole (DDBH),
        \item a rotating gas of gravitons, the so-called Grey Galaxy (GG).        
    \end{itemize}
    Superradiant instabilities are associated with the emission of angular momentum or charge from a rotating or charged black hole. In \cite{Bekenstein:1973mi}, Bekenstein realized that superradiance can, at least at a heuristic level, be derived directly from the laws of black hole mechanics. Here we follow the more recent presentation of \cite{Brito:2015oca}.

\subsection{Thermodynamics for charge superradiance}
    Consider the first law of thermodynamics for a charged, static black hole
    \begin{equation}\label{CSUperfirstlaw}
    \d M =\frac{\kappa}{8\pi}\d A+\mu\,\d Q,
    \end{equation}
    where $M$ is the black hole energy, $\kappa$ the surface gravity, $A$ the area of the event horizon and $\mu$ the chemical potential conjugate to the black hole charge $Q$. Consider the scattering of an incident probe with energy $q_n$ and charge $q_c$. Conservation of charge and energy implies
    \begin{equation}
    \delta Q= \frac{q_c}{q_n}\delta M,
    \end{equation}
    which, upon using \eqref{CSUperfirstlaw}, gives
    \begin{equation}
    \d M = \frac{\kappa}{8\pi} \frac{q_n}{q_n- q_c \mu} \d A.
    \end{equation}
    The area theorem, $\d A\geq0$, then implies that whenever
    \begin{equation}
    \mu>\fr{q_n}{q_c},
    \end{equation}
    the probe can extract energy from the black hole. Note that a probe will respect some BPS bound $q_n\geq q_c$, thus making a black hole with $\mu>1$ unstable.
    
    This general argument acquires a concrete realization in the D3-brane picture proposed in \cite{DDBH} for charge superradiance. The authors consider a probe dual Giant Graviton -- a D3-brane wrapped in the spacetime directions and orbiting along maximal cycles in the internal ones -- in the background of a five-dimensional AdS black hole with equal charges and angular momenta in $AdS_5\times S^5$. By studying the minima of the effective potential governing the probe dynamics, they link the onset of charge superradiant instability to the emission of a charged dual Giant Graviton. We study a generalization to AdS$_5\times T^{1,1}$ and find that the same conclusions  hold in this case. 

    Brane emission can be understood from a thermodynamic perspective by considering a quasi-non-interacting\footnote{The only interaction effect is the reduction of the effective five-form flux $N$ by one unit for every emitted brane.} system composed of a five-dimensional black hole and a small number $m\ll N$ of probe dual Giant Gravitons. The original argument for charge superradiance was presented in Section~3 of \cite{DDBH}. We represent it here and then reshape it to treat angular momentum superradiance. \\
    Even though in this work we will consider black hole solutions of five-dimensional minimal gauged supergravity with one charge and effectively one angular momentum, the thermodynamic argument holds for more general AdS black holes. We, thus, consider a more general five-dimensional AdS black hole carrying energy $E^{BH}$, charges $Q_i^{BH}$ (where $i=1,2,3$ in the five-sphere truncation, while $i=1,2$ in the conifold) and angular momenta $J_j^{BH}$ ($j=1,2$). Each probe carries energy $E^{D3}$ together with charges $Q_i^{D3}$. For dual Giant Gravitons, the BPS relation implies 
    \begin{equation}
        E^{D3}=\sum_i Q_i^{D3}.
    \end{equation}
   Following \cite{DDBH}, we assume all conserved quantities to be positive without loss of generality. Conservation of charge and energy implies that the black hole quantum numbers are reduced by the amounts carried by the emitted probes. Denoting by $N_{eff}$ the effective flux number, $E^{\mathrm{tot}}$ the total energy, $Q_i^{\mathrm{tot}}$ the total charge, and $J_{j}^{\mathrm{tot}}$ the total angular momentum of the black hole-probe brane system, one can write
    \begin{equation}
        \begin{split}
            N_{eff}=N\l1-\fr m N\r,\quad E^{BH}=E^{tot}-E^{D3},\quad Q_i^{BH}=Q_i^{tot}-Q_i^{D3},\quad J_j^{BH}=J_j^{tot}.
        \end{split}
    \end{equation}
    The entropy of the system is given by the black hole entropy
    \begin{equation}
        S_{BH}=N_{eff}^2 \,s_{BH}\l \frac{E^{BH}}{N_{eff}^2},\frac{Q_i^{BH}}{N_{eff}^2},\frac{J_j^{BH}}{N_{eff}^2} \r.
    \end{equation}
    By expanding $S_{BH}$ in $\fr m N$, one obtains   
    \begin{equation}
    \begin{split}
           \frac{S_{BH}}{N^2}=&s^0_{BH}+2 \fr m N \l E^{BH} \fr{\de s_{BH}}{\de E^{BH}} +Q_i^{BH} \fr{\de s_{BH}}{\de Q_i^{BH}}+J_j^{BH} \fr{\de s_{BH}}{\de  J_j^{BH}}-s^0_{BH}\r+\cO\l\fr m N\r^2,\\
           =&s^0_{BH}+2\b \fr m N  \l \frac{E^{BH}}{N^2}-\mu_i\frac{Q_i^{BH}}{N^2}-\Omega_j\frac{J_j^{BH}}{N^2} -\fr{s^0_{BH}}\b\r+\cO\l\fr m N\r^2  ,
    \end{split}
    \end{equation}
    where $\beta$, $\mu_i$ and $\Omega_j$ are identified with the inverse temperature, the electric chemical potentials and the horizon angular velocities through
    \begin{equation}
        \begin{split}
             \b=N^2 \fr{\de s_{BH}}{\de E^{BH}},\quad
            \b\,\mu_i=-N^2 \fr{\de s_{BH}}{\de Q_i^{BH}},\quad \b\,\Omega_j=-N^2 \fr{\de s_{BH}}{\de J_j^{BH}},
        \end{split}
    \end{equation}
    and we defined
    \begin{eqnarray}
         s^0_{BH}=s_{BH}\l \frac{E^{BH}}{N^2},\frac{Q_i^{BH}}{N^2},\frac{J_j^{BH}}{N^2} \r.
    \end{eqnarray}
    The maximization of the leading order entropy gives
    \begin{equation}
        \frac{\de S_{BH}}{\de Q_i^{D3}}=N^2 \frac{\de }{\de Q_i^{D3}}\, s_{BH}\l\frac{E^{tot}-\sum_{\tilde i}Q_{\tilde i}^{D3}}{N^2},\frac{Q_{\tilde i}^{tot}-Q_{\tilde i}^{D3}}{N^2},\frac{J_j^{BH}}{N^2} \r=\b\l\mu_i-1\r=0,
    \end{equation}
    which leads to either $\mu_i=1$ or $Q_i^{D3}=0$. Being $\mu_i$ an increasing function of $Q_i^{BH}$, we can have 
    \begin{itemize}
        \item $Q_i^{D3}=0$ and a vacuum black hole phase if $\mu_i<1\,\,\forall i$
        \item  $\mu_i=1$ for at least one $i$ and a mixed state with $Q_i^{D3}$-charged D3-branes emitted in the $i^{th}$ direction. 
    \end{itemize}
    At next-to-leading order, entropy maximization requires evaluating the Gibbs free energy
    \begin{equation}
    G=\frac{E^{BH}}{N^2}-\mu_i\frac{Q_i^{BH}}{N^2}-\Omega_j\frac{J_j^{BH}}{N^2}-\frac{s^{0}_{BH}}{\beta}.
    \end{equation}
    We specialize to the black hole solution with equal charges $Q^{BH}_{i}=Q$ and angular momenta $J^{BH}_j=J$. Measuring the charges in units of $N$, the Gibbs free energy becomes
    \begin{equation}
    G=E^{BH}-3\mu\,Q^{BH}-2\,\Omega_H\, J^{BH}-\b^{-1} s_0^{BH},
    \end{equation}
    where $\mu$ and $\Omega_H$ are the chemical potentials conjugate to the charge and angular momentum, respectively.\footnote{In the coordinates of \eqref{5dsolution}, $\Omega=-2\Omega_H$. The velocity is indeed minus the sum of the unit angular velocities along the two Cartan directions of $S^3$.\label{note_angvel}}\\
    Since the correction to the entropy is proportional to $m\,G$, the sign of $G$ determines whether the entropy increases or decreases as additional branes are emitted.\\
    Setting $\mu=1$ leads to $$G\leq0.$$
    The entropy is therefore maximized by minimizing the number $m$ of emitted branes, which in this case implies $m=1$. The thermodynamically preferred configuration is thus a quasi-non-interacting mixed state consisting of the central black hole surrounded by a single charged dual Giant Graviton, namely the DDBH configuration discussed in \cite{DDBH,Choi:2025lck}.

\subsection{Thermodynamics for angular momentum superradiance}
    We now want to adapt the above argument for angular momentum superradiance. By exchanging the role of charge and angular momentum in \eqref{CSUperfirstlaw} one can see that energy can be extracted from a spinning black hole via angular momentum superradiance if the horizon angular velocity exceeds one ($\Omega_H>1$). 
    
    As in the charge superradiance case presented above, we consider a quasi-non-interacting system composed of a five-dimensional black hole and a small number $m\ll N$ of emitted objects. We consider a five-dimensional AdS black hole carrying energy $E^{BH}$, charges $Q_i^{BH}$ ($i=1,2,3$ or $i=1,2$, depending on the truncation), and angular momenta $J_j^{BH}$ ($j=1,2$), and probes carrying energy $E^{D3}$ and angular momenta $J_j^{D3}$\begin{comment}\footnote{A proper Giant Graviton will eventually be also electrically charged, while a baryonic probe will have its baryonic charge.}\end{comment}. We write the energy of the probe as 
    \begin{equation}
        E^{D3}=\sum_j J_j^{D3}+\cdots,
    \end{equation}
    where the dots represent the other charges the probe might have due to the embedding in the internal directions, for example baryonic charge. Assume all conserved quantities to be positive. Denoting by $N_{eff}$ the effective flux number, $E^{\mathrm{tot}}$ the total energy, $J_j^{\mathrm{tot}}$ the total angular momenta, and $Q_i^{\mathrm{tot}}$ the total electric charge, one finds
    \begin{equation}
        \begin{split}
            N_{eff}=N\l1-\fr m N\r,\quad E^{BH}=E^{tot}-E^{D3},\quad J_j^{BH}=J_j^{tot}-J_j^{D3},\quad Q_i^{BH}=Q_i^{tot}.
        \end{split}
    \end{equation}
    The entropy of the system is given by the black hole entropy
    \begin{equation}
        S_{BH}=N_{eff}^2 \,s_{BH}\l \frac{E^{BH}}{N_{eff}^2},\frac{Q_i^{BH}}{N_{eff}^2},\frac{J_j^{BH}}{N_{eff}^2} \r.
    \end{equation}
    By expanding $S_{BH}$ in $\fr m N$, one obtains 
    \begin{equation}
    \begin{split}
           \frac{S_{BH}}{N^2}=&s^0_{BH}+2 \fr m N \l E^{BH} \fr{\de s_{BH}}{\de E^{BH}} +Q_i^{BH} \fr{\de s_{BH}}{\de Q_i^{BH}}+J_j^{BH} \fr{\de s_{BH}}{\de  J_j^{BH}}-s^0_{BH}\r+\cO\l\fr m N\r^2,\\
           =&s^0_{BH}+2\b \fr m N  \l \frac{E^{BH}}{N^2}-\mu_i\frac{Q_i^{BH}}{N^2}-\Omega_j\frac{J_j^{BH}}{N^2} -\fr{s^0_{BH}}\b\r+\cO\l\fr m N\r^2  ,
    \end{split}
    \end{equation}
    where $\beta$, $\mu_i$ and $\Omega_j$ are identified with the inverse temperature, the electric chemical potentials and the horizon angular velocities through
    \begin{equation}
        \begin{split}
             \b=N^2 \fr{\de s_{BH}}{\de E^{BH}},\quad
            \b\,\mu_i=-N^2 \fr{\de s_{BH}}{\de Q_i^{BH}},\quad \b\,\Omega_j=-N^2 \fr{\de s_{BH}}{\de J_j^{BH}},
        \end{split}
    \end{equation}
    and we defined
    \begin{eqnarray}
         s^0_{BH}=s_{BH}\l \frac{E^{BH}}{N^2},\frac{Q_i^{BH}}{N^2},\frac{J_j^{BH}}{N^2} \r.
    \end{eqnarray}
    The maximization of the leading order entropy gives
    \begin{equation}
        \frac{\de S_{BH}}{\de J_j^{D3}}=N^2 \frac{\de }{\de J_j^{D3}} s_{BH}\l\frac{E^{tot}-\l \sum_{j}J_j^{D3}+\cdots\r}{N^2},\frac{Q_i^{BH}}{N^2},\frac{J_j^{tot}-J_j^{D3}}{N^2} \r=\b\l\Omega_j-1\r=0,
    \end{equation}
    which leads to either $\Omega_j=1$ or $J_j^{D3}=0$. Being $\Omega_j$ an increasing function of $J_j^{BH}$, we can have 
    \begin{itemize}
        \item $J_j^{D3}=0$ and a vacuum black hole phase if $\Omega_j<1\,\,\forall j=1,2$
        \item  $\Omega_j=1$ for at least one $j$ and a mixed state with D3-branes emitted with $J_j^{D3}$ angular momentum in the $j^{th}$ direction. 
    \end{itemize}
    At next-to-leading order, entropy maximization requires evaluating the Gibbs free energy, which, for a black hole with equal charges and angular momenta reads
    \begin{equation}
    G=E^{BH}-3\mu,Q^{BH}-2\Omega_H J^{BH}-\b^{-1}s_0^{BH},
    \end{equation}
    where $\mu$ and $\Omega_H$ are the chemical potentials conjugate to the charge and angular momentum, respectively.\footnote{See footnote \ref{note_angvel}.}\\
    Since the correction to the entropy is proportional to $m\,G$, the sign of $G$ determines whether the entropy increases or decreases as additional branes are emitted.
    By setting $\Omega_H=1$, one can compute    $$
    G\geq 0.
    $$
    In this case, we have the opposite behavior with respect to charge superradiance: the entropy is maximized for large\footnote{Here ``large'' should still satisfy the condition $m\ll N$.} values of $m$. The thermodynamically favored configuration is therefore a quasi-non-interacting gas of probes orbiting around the central black hole. This is in agreement with the Grey Galaxy picture proposed in \cite{Kim:2023sig,Choi:2025lck}.

    In this work, we take as probes D3-branes wrapped around internal directions. In the maximally symmetric five-sphere truncation, these correspond to Giant Gravitons. In order not to collapse to a point-like graviton, a Giant Graviton rotates at the speed of light along a maximal cycle in the internal directions and is therefore charged. We also allow the Giant Graviton to rotate around the central black hole, thereby giving it angular momentum and stabilizing it outside the event horizon. The presence of charge requires entropy maximization also with respect to the charge. Upon reducing to the universal solution with one charge and one angular momentum, this implies that either the charge vanishes or the system is in the supersymmetric limit, where both chemical potentials are equal to one. The zero-charge case corresponds to shrinking the Giant Graviton to a point-like graviton, which is consistent with the results of Sec. \ref{GGS5} and with the Grey Galaxy picture of \cite{Kim:2023sig}, in which a central black hole is surrounded by a gas of gravitons. In the supersymmetric limit, on the other hand, there is an ambiguity, as the entropy is automatically maximized at leading order. This seems compatible with the picture emerging from Sec. \ref{GGS5}, where a line of degenerate minima connects the Giant Graviton at maximal size with the point-like graviton. 
    
When considering $T^{1,1}$ as the internal space (or, analogously, $Y^{p,q}$), some additional subtleties arise. First, the probe can carry baryonic charge, which may seem to preclude its emission due to charge conservation. However, \cite{Choi:2026faq} proposed a dual field-theory interpretation in which a central black hole surrounded by a baryonic condensate arises as an intermediate configuration along an RG flow from a baryonic black hole in AdS$_5\times T^{1,1}$ to a black hole in AdS$_5\times S^5$. In this picture, the excess baryonic charge is inherited from the UV baryonic black hole, thereby resolving the apparent charge-conservation issue. Alternatively, one may abandon the emission interpretation and the associated thermodynamic picture, and consider other possible interpretations.

A second subtlety is that the wrapped brane also carries R-charge, so the same reasoning applied above to dual Giant Gravitons must be taken into account. For the supersymmetric case we have $\mu =1$ so no problem arises, while away from the susy limit one needs to consider a more general solution, a mixed phase interpolating between the Grey Galaxy and DDBH pictures, with a subset of chemical potentials for charge and angular momentum equal to one. We won't consider this more elaborate case, and our claims on angular momentum superradiance will mostly refer to the supersymmetric configurations.

\section{Dual Dressed Black Holes in AdS$_5\times T^{1,1}$ \label{spacetime_filling}}

We start with Dual Dressed Black Holes (DDBHs): non-interacting bound states of a central
black hole and a dual Giant Graviton, i.e.\ a D3-brane wrapping the $S^3\subset\text{AdS}_5$
and orbiting along a maximal cycle of the internal manifold. The only sizeable interaction is
the reduction of the five-form flux by one unit for each dual Giant Graviton, $N\mapsto N-1$.
DDBHs were constructed in \cite{DDBH} as the endpoint of charge superradiance in
AdS$_5\times S^5$, dual to $\mathcal{N}=4$ SYM. Here we extend the construction to the less
symmetric AdS$_5\times T^{1,1}$ background, dual to the $\mathcal N=1$ Klebanov--Witten
theory \cite{Klebanov:1998hh}.

In AdS$_5\times X^5$ compactifications the charged fields come from modes carrying momentum on
$X^5$, and their number is infinite. In the dual field theory they correspond to half-BPS
single-trace operators: products of the six scalars of $\mathcal N=4$ SYM, or of gauge-invariant
combinations of the bifundamentals $A_iB_j$ in the Klebanov--Witten theory.

We consider a single dual Giant Graviton in the asymptotically AdS$_5\times T^{1,1}$ background
\eqref{DDBHT11metric} and extract its effective potential from the DBI action. At the onset of
charge superradiance, the emission of such a probe is signaled by a stable minimum of the
potential outside the horizon.

Charged AdS black holes were originally expected to be unstable toward the formation of charged
scalar condensates, with hairy black holes as the endpoint. In AdS$_5\times S^5$ it was argued
in \cite{DDBH} that the known hairy solutions are themselves superradiantly unstable, and
hence not the true endpoint. Consistent truncations on $T^{1,1}$ also contain charged scalars
\cite{Cassani:2010na,Bena:2010pr,Halmagyi:2011yd}, so hairy black holes can exist there too, but
no explicit hairy rotating solution is known. We therefore assume,
by analogy with the maximally symmetric case, that hairy black holes are unstable to charge
superradiance.

\subsection{The effective Lagrangian}
        Consider the action for the probe D3-brane,
        \begin{equation}\label{DDBHT11action}
        \begin{split}
        S&=S_{DBI}+S_{WZ},\\
        S&= \frac{1}{(2 \pi)^3 \alpha^{'2} g_s } \left(  -\int d^4y \sqrt{-{\rm det}(P[g_{\mu\nu}])}+  \varepsilon\int P[C_{(4)}]\right), \\
        &= \frac{27\,N}{32\, \pi^2 }\left(  -\int d^4y \sqrt{-{\rm det}(P[g_{\mu\nu}])}+   \varepsilon \int P[C_{(4)}] \right), \\
        \end{split}
        \end{equation}
        where $P[\cdot]$ is the pull-back of the $10$-dimensional metric to the brane world-volume,
        $y$ are the world-volume coordinates, $\varepsilon=\pm1$ for brane or antibrane
        configurations, and \eqref{T11AdSradius} was used from the second to the third line. We set
        $\ve=1$, which in our conventions corresponds to brane configurations.\\
        Using the ten-dimensional uplift \eqref{DDBHT11metric}, we consider the embedding of a
        D3-brane wrapping the $S^3\subset\text{AdS}_5$ and moving along the radial and internal
        directions:
        \begin{equation}\label{DDBHT11embedding}
        \begin{split}
            t=&\tau,\quad r=r(\tau), \quad \psi_a=\psi_0, \quad \phi_a=\phi_0,\quad \theta_a=\theta_0 \\
             \psi=&\psi(\tau),\quad \theta_{i=1,2}=\theta_{i=1,2}(\tau),\quad \phi_{i=1,2}=\phi_{i=1,2}(\tau)\,.
        \end{split}
        \end{equation}
        Here $\psi_a,\theta_a,\phi_a$ are coordinates of AdS$_5$, while $\psi,\theta_i,\phi_i$,
        $i=1,2$, are coordinates on the internal $T^{1,1}$. To lighten the notation we define
        \begin{equation}\label{DDBHT11rescaledN}
            \tilde N \equiv \frac{27}{16} N.
        \end{equation}
        The WZ term reduces to
        \begin{equation}
        \text{$\cal L$}_{WZ}= \tilde N \left( r^4-R^4 \right),
        \end{equation}
        while in the DBI term the square root of the pulled-back metric determinant is
        \begin{equation}
            \sqrt{-{\rm det}(P[g_{\mu\nu}])}=\frac{1}{24} m\,\sqrt{\Delta} \,\sin{\theta_0},
        \end{equation}
        with
        \begin{equation}
            m=2\,r^2 \sqrt{A_{\sigma_3}^2 +b^2},
        \end{equation}
        and where $\Delta$, which encodes the details of the internal manifold, is
        \begin{equation}\label{DDBHT11deltadef}
            \begin{split}
                \Delta= &\frac{\,r^2  W^2 -4 \,\dot r ^2 b^2}{4\,b^2 W} -\frac{1}{4}\left(\frac23\,\dot\theta_i\dot\theta_i -\frac19\cos({2\theta_i})\dot\phi_i^2+ \frac13 \dot\phi_i\dot\phi_i \right) +\frac{\,A_{\sigma_3}^2\cos(2\theta_i)\dot\phi_i^2-\,b^2\dot\phi_i\dot\phi_i}{18\left(A_{\sigma_3}^2+b^2\right)}+\\ &{}-\frac{b^2}3\frac{A_t-A_{\sigma_3 }f}{A_{\sigma_3}^2+b^2}\left( 2  \cos(\theta_i) \dot\phi_i +2\,\dot\psi+3  \left(A_t-A_{\sigma_3 }f\right)\right)+\\
                &-\frac19\frac{b^2}{A_{\sigma_3}^2+b^2}\Big( \cos(\theta_i)\dot\phi_i\dot\psi+\cos(\theta_1)\dot\phi_1\cos(\theta_2)\dot\phi_2+\\
                &+\left( \cos(\theta_2)\dot\phi_2+\dot\psi\right)\left(\cos(\theta_1)\dot\phi_1+\dot\psi\right)\Big).
            \end{split}
        \end{equation}
        After integration over the $S^3$, the effective Lagrangian of the probe is
       \begin{equation}\label{DDBHT11lagrangian}
           \begin{split}
               \text{$\cal L$}_{eff}&=\text{$\cal L$}_{DBI}+ \text{$\cal L$}_{WZ},\\
               &=\tilde N\left(-m\,\sqrt{-\Delta} +  r^4-R^4\right).
           \end{split}
       \end{equation}

\subsubsection{Motion on the squashed $T^{1,1}$}\label{Motion_on_the_squahed_T11}

        Consider the motion of the brane on the $T^{1,1}$. The metric \eqref{DDBHT11metric}
        preserves the $SU(2)_1\times SU(2)_2\times U(1)_{\psi}$ symmetry of $T^{1,1}$, and we want
        to identify equivalent orbits. The Killing vectors are
        \begin{equation}\label{DDBHT11killings}
        \begin{split}
                 \xi^{(1,2)}_{1} &= 
         \sin\phi_{1,2}\,\partial_{\theta_{1,2}}
         + \cot\theta_{1,2}\cos\phi_{1,2}\,\partial_{\phi_{1,2}}
         - \csc\theta_{1,2}\cos\phi_{1,2}\,\partial_{\psi}, \\[4pt]
        \xi^{(1,2)}_2 &=
         -\cos\phi_{1,2}\,\partial_{\theta_{1,2}}
         + \cot\theta_{1,2}\sin\phi_{1,2}\,\partial_{\phi_{1,2}}
         - \csc\theta_{1,2}\sin\phi_{1,2}\,\partial_{\psi}, \\[4pt]
        \xi^{(1,2)}_3 &=\partial_{\phi_{1,2}},\\
        \xi_{\psi}{\,\,\,\,\,\,} &= \partial_{\psi}.
        \end{split}
        \end{equation}
        For each $SU(2)$ there is a vector of charges $Q^{(1,2)}=P^i\xi_i^{(1,2)}$, where $P$ is
        the vector of momenta conjugate to the $T^{1,1}$ coordinates. In the $S^5$ case the
        $SO(6)$ symmetry of the sphere is broken to $U(3)$ by the squashing induced by the
        one-form $A$. Here we have $7$ charges instead of $9$, but none of the symmetry is broken:
        along the internal manifold the one-form $A$ does not depend on the $T^{1,1}$ coordinates
        and merely shifts the fiber coordinate $\psi$ by a constant. We can thus use each $SU(2)$
        to rotate the corresponding charge vector along one axis,
        \begin{equation}
            Q^{(1,2)}_i=(0,0,q^{(1,2)}),
        \end{equation}
        which in our coordinates means $Q^{(1,2)}\propto\partial_{\phi_{1,2}}$. This choice fixes
        $\theta_{1,2}$ and their conjugate momenta in terms of the remaining momenta
        $p_{\f_{1,2}}$ and $p_{\psi}$:
        \begin{equation}\label{DDBHT11motionT11}
            p_{\theta_{1,2}}=0,\quad \theta_{1,2}=\arccos\left(\frac{p_{\psi}}{p_{\phi_{1,2}}}\right).
        \end{equation}
       We are left with three free charges. This seems at odds with the DDBH being rank-2 objects
       for black holes with three equal charges \cite{DDBH}, but $T^{1,1}$ has one more symmetry
       that we have not used yet, the $\mathbb{Z}_2$ exchange of $SU(2)_1$ and $SU(2)_2$:
        \begin{equation}
            \begin{split}
                \theta_1 \longleftrightarrow \theta_2,\quad
                \phi_1\longleftrightarrow\phi_2 .
            \end{split}
        \end{equation}
        As a result the potential depends on $p_{\phi_1}$ and $p_{\phi_2}$ only through a single
        combination.

        \nin
        We define all momenta in units of the rescaled color number $\tilde N$ of
        \eqref{DDBHT11rescaledN},
        \begin{equation}\label{DDBHT11momentadef}
          \tilde N \,  p_{\psi}=\frac{\partial \text{$\cal L$}_{eff}}{\partial\dot\psi }, \quad \tilde N \,  p_{\phi_{1,2}}=\frac{\partial \text{$\cal L$}_{eff}}{\partial\dot\phi_{1,2}},
        \end{equation}
        and solve \eqref{DDBHT11momentadef} for the velocities $\dot \f_{1,2}$ and $\dot \psi$:
        \begin{equation}
        \begin{split}\label{DDBHT11pmomentasol}
             \dot\phi_{1,2}&=3\, p_{\phi _{1,2}}\frac{1}{\sqrt{K}}\sqrt{\frac{W^{2}\,r^{2} - 4 b^{2}\, \dot r^{2}}{W}},\\
             \dot\psi=&\frac{3}{K}\Bigg((A_{\s_3} f - A_t)\Big(-3\big(3A_{\s_3}^{2}p_{\psi}^{2}
             + b^{2}\big(2(p_{\phi_1}^{2}+p_{\phi_2}^{2})-p_{\psi}^{2}\big)\big)
             +4b^{2}\big(A_{\s_3}^{2}+b^{2}\big)r^{4}\Big)+\\
             &\quad+\; \frac{(3A_{\s_3}^{2}-b^{2})\,p_{\psi}}{2\,b^{2}}\,
            \sqrt{\frac{\,K\big(W^{2}r^{2}-4b^{2}\dot r^{2}\big)\,}{W}}\Bigg),\\
            K&= 9A_{\sigma_3}^{2}p_{\psi}^{2}
            + b^{2}\big(6(p_{\phi_1}^{2}+p_{\phi_2}^{2}) - 3p_{\psi}^{2}\big)
            + 4b^{2}\big(A_{\sigma_3}^{2}+b^{2}\big)\,r^{4}.
        \end{split}
        \end{equation}

          \begin{figure}[h!]
            \centering
            \includegraphics[width=\linewidth]{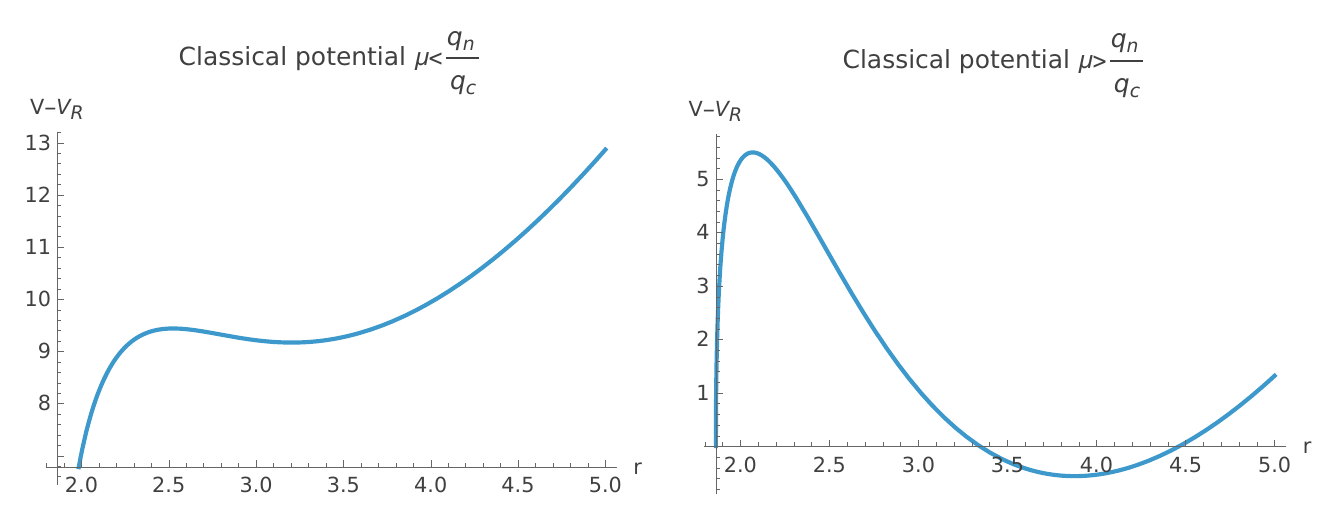}
            \caption{Energy of the probe in units of $\tilde N$ at distance $r$ minus that at the
            horizon, $V_{cl}(r)-V_{cl}(R)$. Here $p=19.236,\ j=0.12318,\ q=6 $, corresponding to a
            horizon radius $R\simeq1.8$ and a chemical potential $\mu\simeq 1.68$. On the left
            $q_c = 1,\ q_n = 10$, on the right $q_c = 9,\ q_n = 10$. For $\mu>\fr{q_n}{q_c}$ the
            minimum lies below $V_{cl}(R)$ and is stable: the black hole can lower its energy by
            emitting the probe, signaling the superradiant instability. For $\mu<\fr{q_n}{q_c}$ the
            minimum lies above $V_{cl}(R)$ and is only metastable. This reproduces Fig.~6 of \cite{DDBH}.}
            \label{fig:vcl}
        \end{figure}

\subsection{The effective potential}

    The effective $r$-dependent potential follows from the Lagrangian
    \eqref{DDBHT11lagrangian} by a Legendre transform, using \eqref{DDBHT11pmomentasol}:
    \begin{equation}
        \label{DDBHT11clpot}
        \begin{split}
             V_{\rm cl}(r)=& \tilde N\,\l p_{\psi}\dot\psi +p_{\phi_1}\dot \f_1 + p_{\f_2} \dot\f_2 \r -\text{$\cal L$}_{eff} |_{\dot r=0},\\
            =& \tilde N\left(\frac{r\,\sqrt{W}}{2b^{2}}\sqrt{9A_{\sigma_3}^{2}p_{\psi}^{2}+b^{2}\!\left(m^2+6p_{\phi_1}^{2}+6p_{\phi_2}^{2}-3p_{\psi}^{2}\right)}\right.\,+\\
            &\,-3p_{\psi}\left(A_t-A_{\sigma_3}f\right)-r^{4}+R^{4}\Biggr).
        \end{split}
        \end{equation}
    As anticipated, the $\mathbb{Z}_2$ symmetry $1\longleftrightarrow2$ makes $V_{\rm cl}$ depend
    only on $p_{\f_1}^2+p_{\f_2}^2$. The problem is thus reduced to an effectively
    one-dimensional one in $r$, in which, at fixed black hole, the probe is characterized by two
    parameters, $p_\f$ and $p_{\psi}$.

    The coefficient of $A_t$ is the brane charge in units of $\tilde N$,
    \begin{equation}
        q_c=-3\,p_{\psi}\in\mathbb{R},
    \end{equation}
    and we further define
    \begin{equation}
        q_n^2=6\,\left(p_{\phi_1}^2+p_{\phi_2}^2\right)-3\,p_{\psi}^2 .
    \end{equation}
    Up to the overall factor $\tilde N$, \eqref{DDBHT11clpot} then coincides with the dual-giant
    potential of the maximally symmetric case (eq.~(4.24) of \cite{DDBH}):
    \begin{equation}\label{DDBHpot}
         \begin{split}
             V_{\rm cl}(r)=\tilde N\Bigg(\frac{ r \sqrt{W}}{2 b^2} \sqrt{q_c{}^2 A_{\sigma _3}^2+b^2 \left(m^2+q_n{}^2\right)}+q_c \left(A_t-f A_{\sigma _3}\right)-r^4+R^4\Bigg).
        \end{split}
    \end{equation}
    The constraint from \eqref{DDBHT11motionT11}, $|p_{\f_{1,2}}|\geq|p_\psi|$ (required for the
    arccosine to be defined), implies
    \begin{equation}
        q_n^2-q_c^2=6\left(p_{\phi_1}^2+p_{\phi_2}^2\right)-12\,p_\psi^2\;\ge\;0,
        \qquad\text{i.e.}\qquad |q_n|\ge|q_c| ,
    \end{equation}
    in agreement with \cite{DDBH}. Our analysis is therefore valid for $|p_{\f_{1,2}}|\geq|p_\psi|$.

    Fig.~\ref{fig:vcl} shows the potential for $\mu<\frac{q_n}{q_c}$ and $\mu>\frac{q_n}{q_c}$:
    the minimum is metastable in the first case and stable in the second. For $q_n=q_c$ these
    are the conditions for charge superradiance, so we expect AdS$_5\times T^{1,1}$ black holes
    with $\mu>1$, unstable under charge superradiance, to decay into a Dual Dressed Black Hole
    phase. In general, $V_{\rm cl}$ is a complicated function of $r$, and can have zero, one or two
    minima outside the horizon depending on the black hole and on the charges. These can be
    scanned by defining $\alpha= q_c/q_n\in (-1,1)$ and varying the charges. We restrict to a few
    physically relevant limits.

    Since \eqref{DDBHpot} is identical in form to the potential of \cite{DDBH}, the analysis of
    its limits carries over unchanged (see \cite{DDBH} and references therein). In the rest of this
    subsection we report the salient points of their analysis \cite{DDBH} and the results we need, adapted to our notation. Energies and
    charges are measured in units of $\tilde N$, which we set to one.

\subsubsection{Analysis of the effective potential}

\begin{figure}[h!]
            \centering
            \includegraphics[width=0.5\linewidth]{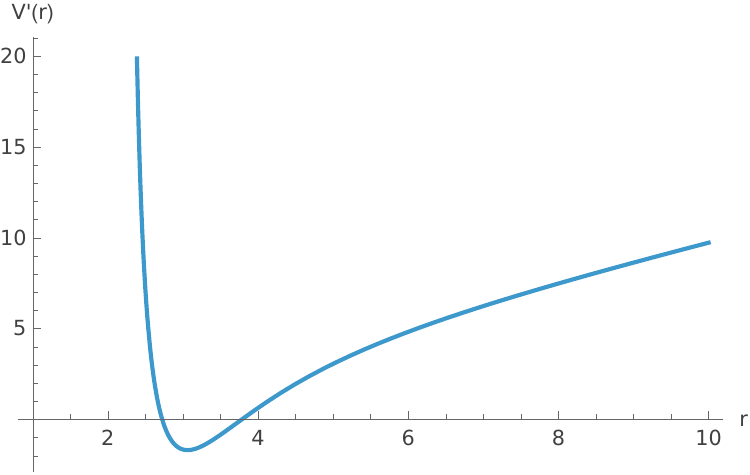}
            \caption{Behavior of the derivative of the effective potential with respect to $r$, $V_{cl}'(r)$, in the neighborhood of the horizon. The values of the parameters used for this plot are $R = 1,\,\, q=6,\,\, j=0.12, \,\,q_n=10,\,\, q_c=10 $. }
            \label{fig:vcl'}
        \end{figure}

         At $r=R$ the potential in the gauge \eqref{5dsolution} reduces to
        \begin{equation}\label{DDBHS5vclR}
            V_{cl}(R)= \mu\, q_c,
        \end{equation}
        with the chemical potential $\mu$ defined in \eqref{5dsolchemicalpot}; in the gauge
        regular at the horizon, $A_t\mapsto A_t-\mu$, it vanishes. In both gauges we checked
        numerically that $V_{cl}'(R)>0$ for all $\mu$ (a prime denotes $d/dr$), see
        Fig.~\ref{fig:vcl'}.
        
        For $r\to\infty$ the potential grows quadratically,
        \begin{equation}
            V_{cl}(r) =  \frac{r^2}{2}+ {\cal O}(1).
        \end{equation}
        When $q_n$ is also large, with $q_n/r^2$ fixed, $V_{cl}$ approaches the pure AdS
        potential
        \begin{equation}\label{DDBHS5potenialAdS}
            V_{\rm cl}(r)= \sqrt{\left(1+ r^2\right)\left(q_{n}^2 + r^6\right)} -r^4+{\cal O}(1),
        \end{equation}
        whose minimum sits at $r = \sqrt{q_ n }$ with $E= q_n$, the BPS value in AdS.

        Corrections to this can be computed by expanding in inverse charges,
        $1/q_n\sim 1/q_c \sim 1/r^2\sim \cO (\epsilon)$ with $\epsilon \to 0$:
        \begin{equation}
            V_{cl}(r) = \frac{q_n^2+r^4}{2r^2}\frac{1
            }{\epsilon}+\frac{r^8\left(8p\left( j^2-1\right) -1+8R^4 \right) +q_n^2\left(2r^4-q_n^2\right)+8q\, r^6\left(q_c+r^2 \right)}{8r^8} +{\cal O}\left( \epsilon \right).
        \end{equation}
        Here $\epsilon$ is a bookkeeping parameter, set to one at the end. The minimum and its
        energy are
        \begin{equation}
            r_{min} = \sqrt{q_n}+ \frac{q\, q_c}{2\,q_n^{3/2}}\epsilon,
            \qquad
            E \equiv V_{\rm cl}(r_{\rm min})= \frac{q_n}{\epsilon} +\left(\left(j^2-1\right) p+\frac{q \,q_{c}}{q_n}+q+R^4\right)+{\cal O}\left( \epsilon \right).
        \end{equation}
        Comparing $E$ with the horizon value \eqref{DDBHS5vclR}:
        \begin{itemize}
        \item if $\mu < \frac{q_n}{q_c}$, then $V_{\rm cl}(r_{\rm min}) > V_{\rm cl}(R)$: the
        minimum is metastable, and the probe can lower its energy by tunneling into the black hole;
        \item if $\mu > \frac{q_n}{q_c}$, then $V_{\rm cl}(r_{\rm min}) < V_{\rm cl}(R)$: the
        minimum is stable.
        \end{itemize}
        This result was already found for the $S^5$ in \cite{Henriksson:2019zph,DDBH} and indeed it agrees with the thermodynamic expectation also for $T^{1,1}$. The threshold is reached first for $q_n=q_c$, since this maximizes the charge-to-energy
        ratio at leading order. For $q_n=q_c$ and $\epsilon=1$,
        \begin{equation}
            E=  q_n + p\left(j^2 -1\right) + 2 q + R^4 .
        \end{equation}
        As observed in \cite{DDBH}, the correction $p\left(j^2 -1\right) + 2 q + R^4$ is positive
        for black holes with $\mu =1$, vanishing only in the supersymmetric case; we confirmed this
        numerically for our parameter range. The charge-to-energy ratio is therefore largest at
        large $q_n=q_c$, so a black hole with $\mu>1$ lowers its free energy by emitting a single
        large dual Giant Graviton with $q_c=q_n$, far from the horizon. This makes the
        quasi-non-interacting picture consistent.

\subsubsection{The supersymmetric limit}
        The central black hole becomes the supersymmetric Gutowski--Reall solution
        \cite{Gutowski:2004ez} upon imposing \eqref{5dGRsolsusycond}. A supersymmetric probe obeys
        the BPS bound $E=q_c$. Following the discussion above, we take $q_n = \left|q_c\right|$,
        which has the smallest mass-to-charge ratio and is thus the most likely to saturate the
        bound. The potential then reads
        \begin{equation}\label{DDBHS5GRgeneralpot}
        \begin{split}
            V_{cl}(r) &= \frac{1}{4 r^6 + 
                4 r^2 R^6 - R^8}\Bigg( 4 q_c r^4 R^2-4 r^{10} + 2 q_c r^4 R^4 +\\&+ 4 r^6 R^4 - 
                4 r^6 R^6 + q_c R^8 + r^4 R^8 + 4 r^2 R^{10} - R^{12}+ \\&+ 
                2 r^5\sqrt{\frac{(r^2 - R^2)^2 (1 + r^2 + 2 R^2) (r^4 + R^6) (4 q_c^2 + 
                4 r^6 + 4 r^2 R^6 - R^8)}{r^6}}\Bigg).
        \end{split}
        \end{equation}
        One can manipulate the expression like done in Sec. 4.7 of \cite{DDBH}, and see that for $r>R$ the denominator of \eqref{DDBHS5GRgeneralpot} is positive and $V_{cl}\ge q_c$. The bound is saturated only when the expression under the square root is zero;
        such a point is a global (hence local) minimum with $V_{cl}=q_c$, i.e.\ a supersymmetric
        minimum. It exists whenever
        \begin{equation}\label{DDBHT11susymincond}
            q_{c}=F(r^2), \quad F(r^2)=\frac{2 r^4-r^2 R^4+R^6}{2 \left(r^2+R^2\right)},
        \end{equation}
        i.e.\ at
        \begin{equation}
            r^2 = \frac{1}{4}\left(2 q_c + R^4 \pm \sqrt{4 q_c^2 + R^6 (R^2-8) + 4 q_c R^2 (4 + R^2)}\right),
        \end{equation}
        provided $ q_c\frac{2(r^2- R^2)(2r^4 - r^2R^4 + R^6)}{4 r^6 +  4 r^2 R^6 - R^8} + r^4 - R^4>0$ there. For extremal supersymmetric black holes, the momenta
        \eqref{DDBHT11momentadef} evaluated at\footnote{ Notice that in inverting \eqref{DDBHT11momentadef} to derive \eqref{DDBHT11pmomentasol} we have made the implicit assumption that $q_n \neq q_c$. If we release this assumption we cannot use directly  \eqref{DDBHT11pmomentasol}, but we need to start from scracth and manipulate \eqref{DDBHT11momentadef}.} $\dot\f_1=\dot\f_2=\dot\psi=-1$ reproduce
        \eqref{DDBHT11susymincond}, as expected from the $\kappa$-symmetry analysis of
        Appendix~\ref{ksymmetry}.

\begin{figure}[h!]
            \centering
            \begin{subfigure}[b]{0.32\linewidth}
                \centering
                \includegraphics[width=1\linewidth]{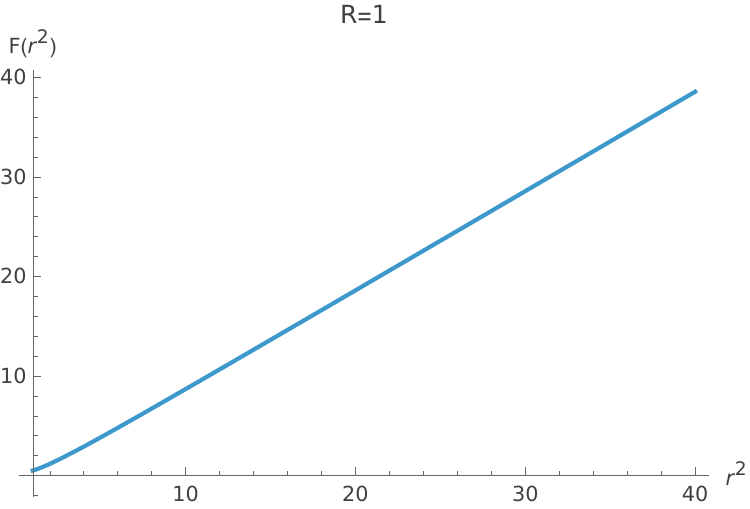}
                \caption{}
            \end{subfigure}
            \begin{subfigure}[b]{0.32\linewidth}
                \centering
                \includegraphics[width=1\linewidth]{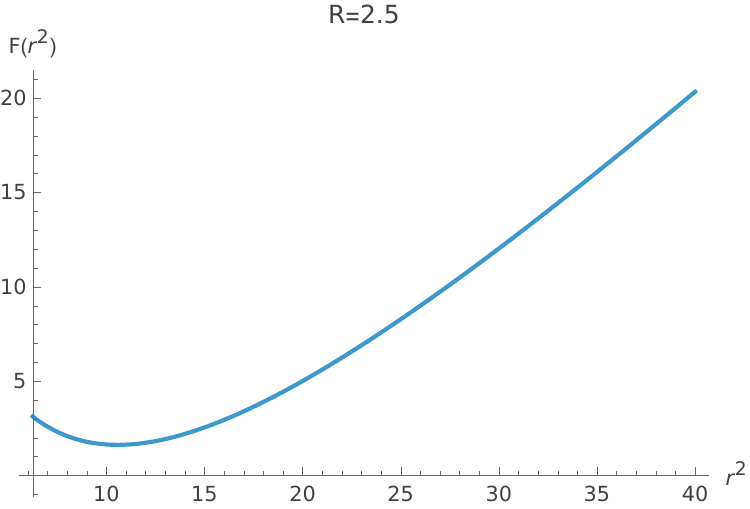}
                \caption{}
            \end{subfigure}
            \begin{subfigure}[b]{0.32\linewidth}
                \centering
                \includegraphics[width=1\linewidth]{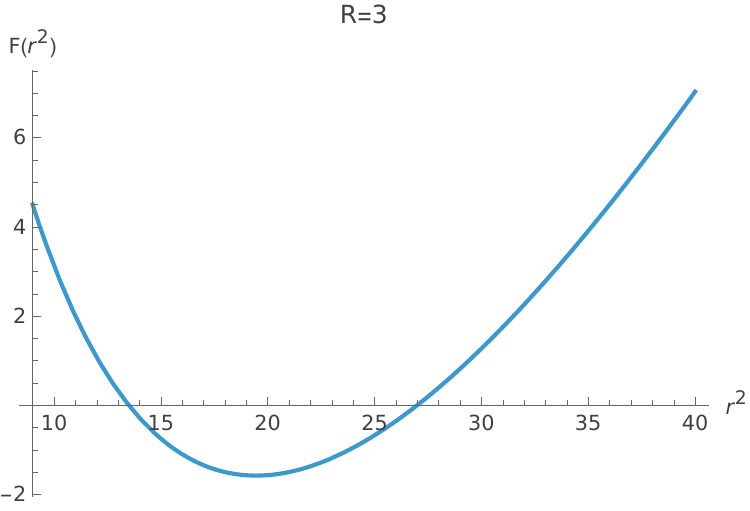}
                \caption{}
            \end{subfigure}
            \caption{The three qualitative behaviors of $F(r^2)$, indicating the existence of zero, one or two supersymmetric minima depending on the horizon position and on the probe charge $q_c$.}
            \label{fig:susymin}
        \end{figure}

\begin{figure}[h!]
    \centering
    
    % Row 1
    \begin{subfigure}[b]{0.495\linewidth}
        \centering
        \includegraphics[width=\linewidth]{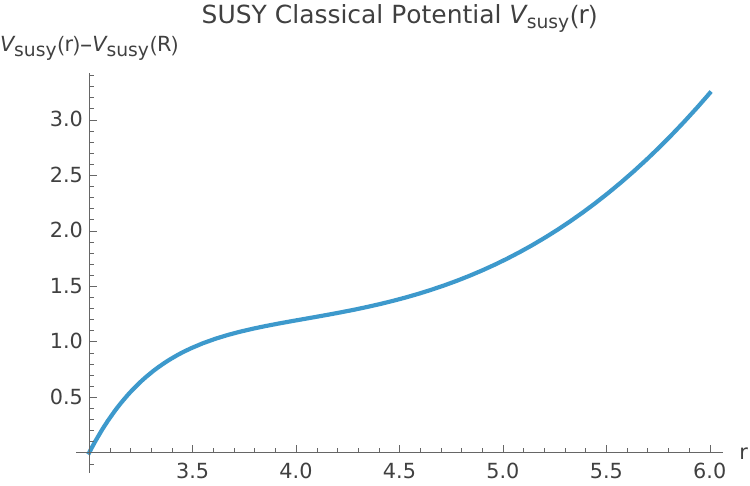}
        \caption{$q_c=-15$.}
        \label{fig:sub1}
    \end{subfigure}
    \hfill
    \begin{subfigure}[b]{0.495\linewidth}
        \centering
        \includegraphics[width=\linewidth]{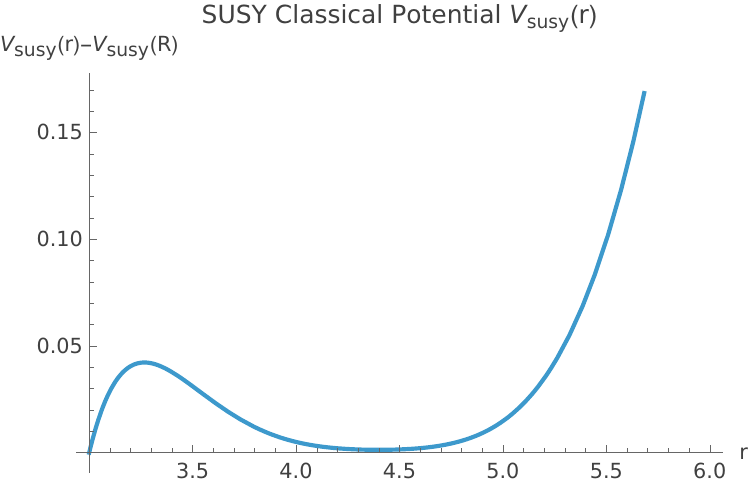}
        \caption{$q_c=-2.$}
        \label{fig:sub2}
    \end{subfigure}

    % Row 2
    \begin{subfigure}[b]{0.495\linewidth}
        \centering
        \includegraphics[width=1\linewidth]{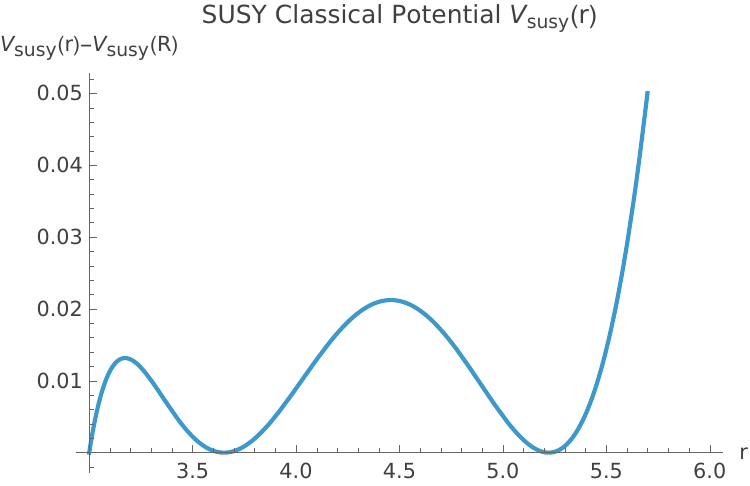}
        \caption{$q_c=0.1$.}
        \label{fig:sub3}
    \end{subfigure}
    \hfill
    \begin{subfigure}[b]{0.495\linewidth}
        \centering
        \includegraphics[width=1\linewidth]{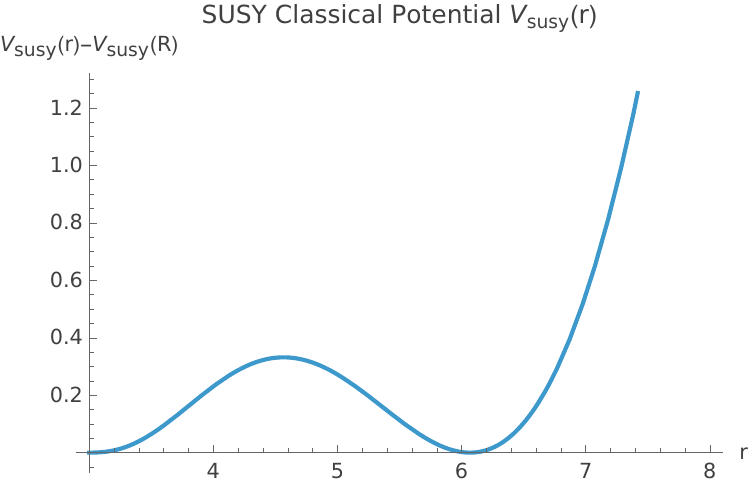}
        \caption{$q_c=5$.}
        \label{fig:sub4}
    \end{subfigure}

    \caption{Energy of a dual Giant Graviton around a supersymmetric black hole with horizon radius $R=3$, for different values of the brane charge $q_c$.}
    \label{fig:DDBHsusypot}
\end{figure}

        The number of supersymmetric minima depends on $R$ and $q_c$, as illustrated in
        Fig.~\ref{fig:susymin}. The most interesting
        situation for us is panel (d) of Fig.~\ref{fig:DDBHsusypot}: there the probe sits far from
        the horizon, as required to treat the black hole and the probe as a weakly interacting
        mixture, whereas in panel (c) it is much closer to the horizon.

        Fig.~\ref{fig:DDBHsusypot} shows the case $R=3$, which has the richest landscape (see
        Fig.~\ref{fig:susymin}). For large negative charge there is no minimum. As the charge
        increases, a quartic minimum (where the second derivative also vanishes) appears and splits
        into two distinct minima, which gradually separate until one reaches the horizon. At large
        positive charge only a single, marginally stable BPS minimum remains, far from the black
        hole.

\section{Wrapped D3-branes on $T^{1,1}$ cycles}\label{GGT11}
        One of the features that makes the internal manifold $T^{1,1}$ particularly interesting is the presence of non-contractible cycles. In the DDBH setup, however, this topological distinction with respect to $S^5$ does not play an essential role, since the D3-branes wrap directions in AdS$_5$ and behave as pointlike objects on the internal manifold. As a consequence, the dynamics effectively reduces to the spherical case, allowing us to extend the same physical picture to this different background.

        We now turn to a genuinely new regime, in which a D$3$-brane wraps the non-contractible cycles of $T^{1,1}$ and the non-trivial topology of the internal space becomes dynamically relevant. In the dual field theory, these configurations correspond to baryon-like operators of the form $A^N$ and $B^N$, obtained by fully antisymmetrizing the color indices of the bifundamental fields and having conformal dimension $\frac{3}{4}N$. The fields $A^{\alpha}{i,\beta}$ carry index $\alpha$ in the $N$ of $SU(N)_1$ and $\beta$ in the $\bar N$ of $SU(N)_2$. The dibaryon is constructed \cite{Gubser:1998fp}:
        \begin{equation}
            {\cal B}_{1l}=\epsilon_{\alpha_1,...,\alpha_N}\epsilon^{\beta_1,...,\beta_N}D_l^{k_1,...,k_N}\prod_{i=1}^{N} A^{\alpha_i}_{k_i\,\beta_i},
        \end{equation}
        where we denote by $D_l^{k_1,...,k_N}$ the completely symmetric $SU(2)$ Clebsch-Gordon coefficient that combines $N$ doublets into the $(N+1)$ representation of $SU(2)$. As a result, ${\cal B}_{1l}$ carries $SU(2)\times SU(2)$ quantum numbers $(N+1,1)$. Analogously, it is possible to construct dibaryons in the $(1,N+1)$ representation. The holographic dual are D3-branes wrapping an $S^3\subset T^{1,1}$ and localized at constant $(\theta_1,\phi_1)$ on one of the $S^2$ factors.

        More general embeddings have been studied, for example in \cite{Henriksson:2019ifu}, in which the D3-brane wraps the two $S^2$ factors with integer winding numbers $m_1$ and $m_2$. These winding numbers determine the baryonic charge and the R-charge of the dual operator. The corresponding field theory operators are expected to be of the form $\left(A^{|m_1|}B^{|m_2|}\right)^N$. They have conformal dimension $\frac{3}{4}(|m_1|+|m_2|)N$ and their baryon number is $(|m_1|-|m_2|)N$. We will consider this more general case.

\subsubsection{Warm-up case: pure AdS$_5\times T^{1,1}$}

    Consider a probe D3-brane in pure AdS$_5\times  T^{1,1}$. Related work was presented in \cite{Hamilton:2010sv,Canoura:2005uz}. The metric is
        \begin{equation}\label{D3pureAdsT11}
        \begin{split}
            ds^2&=ds_{\text{AdS}_5}^2+ds^2_{T^{1,1}},\\
            ds_{\text{AdS}_5}^2&=-\l1+r^2\r dt^2+\frac{dr^2}{1+r^2}+\fr{r^2}{4}\l d\theta_a^2 +d\f_a^2+d\psi_a^2+2\cos\theta_a d\f_a d\psi_a\r,\\
            ds_{T^{1,1}}^2&=\frac{1}{6} \sum_{i=1,2}(d \theta_i^2 + \sin^2 \theta_i d\phi_i^2) 
                        + \frac{1}{9} \left(d \psi +\sum_{i=1,2} \cos \theta_id\phi_i  \right)^2 .
        \end{split}
        \end{equation}
    The coordinate range is $t\in \mathbb{R},\,\,r\in[0,\infty),\,\,\theta_a\in[0,\pi],\,\,\f_a\in[0,2\pi),\,\,\psi_a\in[0,4\pi),\,\,\theta_{1,2}\in[0,\pi],\,\,\f_{1,2}\in[0,2\pi),\,\,\psi\in[0,4\pi)$.
    We adopt the embedding presented in \cite{Henriksson:2019ifu}
     \begin{equation}
        \begin{split}
        t=\tau, \quad r,\psi_a,\phi_a,\theta_a=const,\quad         \theta_{i=1,2}=2\,\arctan\left(c_i\,\zeta^{m_i}\right),\quad \phi_{i=1,2}=m_i\,\beta,\quad \psi=\gamma.
    \end{split}
    \end{equation}
    The brane action is \cite{Henriksson:2019ifu}
    \begin{equation}
         \begin{split} 
        S&= -T_3\int d^4\s \sqrt{-{\rm det}(P[g_{\mu\nu}])}+ T_3   \int P[C_{(4)}], \\
        \end{split}
    \end{equation}
    where $T_3$ is the brane tension. Since we are only interested in the qualitative behavior, we will not keep track of the overall normalization. In this background one finds $P[C_{(4)}]=0$, and the effective Lagrangian reduces to
    \begin{equation}
    \mathcal{L}_{\rm eff}\sim -T_3(m_1+m_2)\sqrt{1+r^2}.
    \end{equation}
    As expected, the energy has a global minimum at the center of AdS$_5$, $r=0$, where the wrapped brane can sit in a stable configuration. In contrast, in pure AdS$_5\times S^5$ (see Sec. \ref{GGS5}), without allowing for motion in the internal manifold, the brane shrinks to zero size. In the $T^{1,1}$ case, instead, the brane is stabilized by the topology of the manifold, and by wrapping non-contractible cycles, it remains stable even when sits still at the center of AdS$_5$. As we are going to see below, an analogous phenomenon occurs when considering a central black hole. \\

\subsection{The effective Lagrangian}

    We now consider a probe \D3-brane in the black hole background \eqref{DDBHT11metric}. The action for the probe brane is
     \begin{equation}\label{GGT11action}
        \begin{split} 
        S&=S_{DBI}+S_{WZ},\\
        S&= \frac{1}{(2 \pi)^3 \alpha^{'2} g_s } \left(  -\int d^4\s \sqrt{-{\rm det}(P[g_{\mu\nu}])}+  \int P[C_{(4)}] \right), \\
        &= \frac{27\,N}{32\, \pi^2 }\left(  -\int d^4\s \sqrt{-{\rm det}(P[g_{\mu\nu}])}+    \int P[C_{(4)}]\right), \\
        \end{split}
    \end{equation}
     where we used \eqref{T11AdSradius} from the second to the third line. Even in this case it results useful to introduce the rescaled color number 
     \begin{equation}
         \tilde N=\fr{27}{16} N.
     \end{equation}
     We consider the embedding presented in Sec. 5 of \cite{Henriksson:2019ifu} for the black hole background in the coordinates \eqref{5dsolutionshifted}, which, in the supersymmetric limit, reduces to \eqref{5dGRsolution}. We show in Appendix \ref{ksymmetry} that, for the Gutowski-Reall black hole \eqref{5dGRsolution}, the following embedding is supersymmetric for $\dot\psi_a=\dot\phi_a=\dot\theta_a=0$:\begin{comment}
         
    \\  \textcolor{red}{here we need to change the metric, it's weird to parameterize $\psi_a$ this way}
     \end{comment} 
    \begin{equation}\label{GGT11embedding}
    \begin{split}
        &t=\tau, \quad r=const,\quad ,\psi_a=\psi_a(\tau),\quad \phi_a=\phi_a(\tau),\quad \theta_a=\theta_a(\tau)\\
            &\theta_{i=1,2}=2\,\arctan\left(c_i\,\zeta^{m_i}\right),\quad \phi_{i=1,2}=m_i\,\beta,\quad \psi=\gamma,
    \end{split}
    \end{equation}
    for $c_{1,2}$ generic constants and $m_{1,2}$ winding numbers.
  
    To simplify the expressions, we define 
    \begin{equation}
         \Xi = \sum_{i=1,2} \frac{c_i^2 m_i^2 \zeta^{2m_i-1}}{\left(1+c_i^2\zeta^{2m_i}\right)^2}.
    \end{equation} 
    The pullback of the Ramond-Ramond four-form potential \eqref{fourfo} gives
    \begin{equation}
        P[C_{(4)}]=-\frac29\l A_t+\l2+\dot\psi_a+\cos\theta_a\dot \f_a\r A_{\s_3}\r\,\Xi \,d\tau\wedge d\zeta\wedge d\beta\wedge d\gamma,
    \end{equation}
   while the square root of the determinant of the pullback metric is
    \begin{equation}
    \begin{split}
        \sqrt{-\det (P[g_{\mu\nu}])}=&\frac{2}{9}\Xi\left(\fr{r^2 W}{4 b^2}
    -b^2 (2 + f)^2-\fr{r^2}{4}\dot{\theta}_a^{\,2}
    -\fr18 \l
    4 b^2+r^2+(4 b^2-r^2)\cos 2\theta_a
    \r\dot{\phi}_a^{\,2}+\right.\\
    &\left.
    -2 b^2 (2+f)\dot{\psi}_a-b^2\dot{\psi}_a^{\,2}
    -2 b^2\cos\theta_a\,\dot{\phi}_a
    \l2+f+\dot{\psi}_a\r
    \right)^{\fr12},
    \end{split}
    \end{equation}
    Integration over $\beta$ and $\gamma$ produces a factor $8\pi^2$, while integration over $\zeta$ gives
    \begin{equation}
        \int_0^{\infty}\Xi\,d\zeta=\frac{|m_1|+|m_2|}{2}.
    \end{equation}
    The resulting Lagrangian is
    \begin{equation}
    \begin{split}\label{GGT11lagrangian}
        \text{$\cal L$}_{eff}=& - \frac89\tilde N \left(|m_1|+|m_2|\right)\left(\sqrt{\bar\Delta}+  A_t+\l2+\dot\psi_a\r A_{\s_3}\right),
    \end{split}
    \end{equation}
    for
    \begin{equation}
        \begin{split}
            \bar\Delta=&\fr{r^2 W}{4 b^2}
    -b^2 (2 + f)^2-\fr{r^2}{4}\dot{\theta}_a^{\,2}
    -\fr18 \l
    4 b^2+r^2+(4 b^2-r^2)\cos \l2\theta_a\r
    \r\dot{\phi}_a^{\,2}+\\
    &
    -2 b^2 (2+f)\dot{\psi}_a-b^2\dot{\psi}_a^{\,2}
    -2 b^2\cos\theta_a\,\dot{\phi}_a
    \l2+f+\dot{\psi}_a\r.
        \end{split}
    \end{equation}
    In the following, we drop the factor $        \fr89\tilde N\l m_1+m_2\r=1$    to ease notation.    Since the black hole solution with equal angular momenta has an $SU(2)$ symmetry, we use it to simplify the motion as proposed in Section \ref{Motion_on_the_squahed_T11} by setting 
    \begin{equation}
        p_{\theta_a}=0,\quad \theta_a=\arccos\l \fr{p_{\psi_a}}{p_{\f_a}}\r,
    \end{equation}
    where $p_{\text{AdS-angle}}$ is the momentum associated with the respective angular variable in AdS$_5$. The Lagrangian simplifies to
    \begin{equation}
    \begin{split}
        \cL_{\rm eff}=&-\left( A_t +\l2+\fr{p_{\psi_a}}{p_{\phi_a}}\dot\phi_a+\dot\psi_a\r A_{\s_3}+\sqrt{\Delta}\right),
    \end{split}
    \end{equation}
    where
    \begin{equation}
    \begin{split}
        \Delta=&\fr{r^2W}{4b^2}-b^2\l2+f\r^2+\l\l \fr{r^2}4-b^2\r \fr{p_{\psi_a}^2}{p_{\phi_a}^2}-\fr{r^2}4 p_{\phi_a}^2\r\dot\phi_a^2+\\
    &-
    2b^2\dot\phi_a
    \l2+f+\dot\psi_a\r
    \fr{p_{\psi_a}}{p_{\phi_a}}
    -b^2\dot\psi_a
    \l4+2f+\dot\psi_a\r.
    \end{split}
    \end{equation}
    We define the conserved momenta
    \begin{eqnarray}\label{GGT11momdef}
    \begin{split}
         p_{\psi_a}=&\frac{\de\cL_{eff}}{\de\dot\psi_a}
         =\frac{b^2\l 2+f+\dot\psi_a+\fr{p_{\psi_a}}{p_{\f_a}}\dot\f_a\r}{\sqrt{\Delta}}- A_{\s_3},\\
         p_{\f_a}=&\frac{\de\cL_{eff}}{\de\dot\f_a}
         =\frac{b^2\fr{p_{\psi_a}}{p_{\f_a}}\l 2+f+\dot\psi_a+\fr{p_{\psi_a}}{p_{\f_a}}\dot\f_a\r+\fr{r^2}4\fr{p_{\f_a}^2-p_{\psi_a}^2}{p_{\f_a}^2}\dot\f_a}{\sqrt{\Delta}}-\fr{p_{\psi_a}}{p_{\f_a}} A_{\s_3}.   \end{split}
    \end{eqnarray}
    In order to solve for $\dot\psi_a$ and $\dot \f_a$, we consider the equation obtained from taking the following linear combination: 
    \begin{equation}
        \fr{\text{II}\fr{p_{\f_a}}{p_{\psi_a}} - \text{I}}{\text{I}+ A_{\s_3}},
    \end{equation}
    where I and II are respectively the first and second equations of \eqref{GGT11momdef}, solve for $\dot\psi_a$ and reinsert in II. We find
    \begin{equation}
    \begin{split}
        \dot\psi_a        =&\frac{ \l p_{\psi_a}+ A_{\s_3}\r r^2-4\,b^2p_{\psi_a} }{2\,b^2}\sqrt{\frac{W}{ \l p_{\psi_a}+ A_{\s_3}\r^2r^2+b^2\l 4\l p_{\f_a}^2-p_{\psi_a}^2\r+r^2 \r}}-\l 2+f\r,\\
        \dot\f_a        =&2p_{\f_a}\sqrt{\frac{W}{ \l p_{\psi_a}+ A_{\s_3}\r^2r^2+b^2\l 4\l p_{\f_a}^2-p_{\psi_a}^2\r+r^2 \r}}.    
      \end{split}\label{dotintermsofp}
    \end{equation}

\subsection{The classical potential}
\begin{figure}[h!]
    \centering
    \begin{subfigure}{0.48\textwidth}
        \centering
        \includegraphics[width=\linewidth]{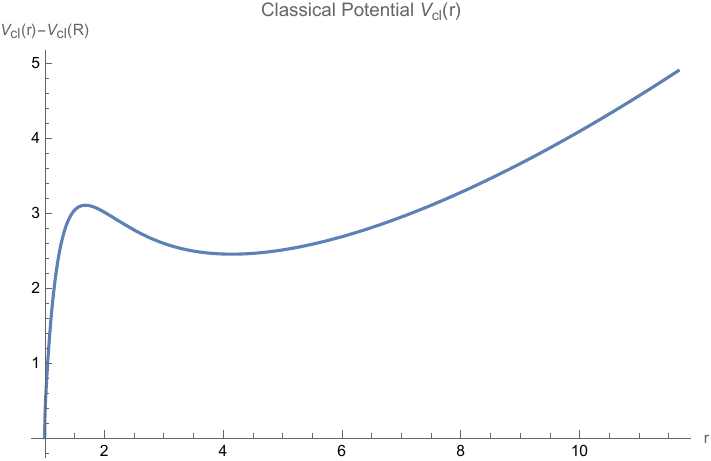}
    \end{subfigure}
    \hfill
    \begin{subfigure}{0.48\textwidth}
        \centering
        \includegraphics[width=\linewidth]{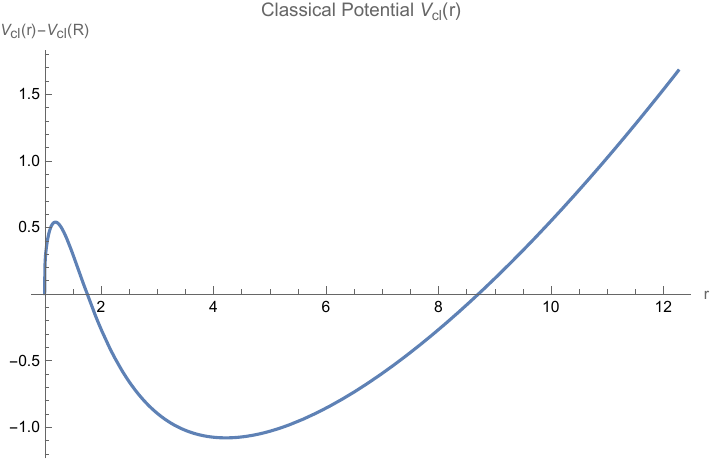}
    \end{subfigure}
    \caption{Energy of the probe at distance $r$ minus that at the horizon $V_{cl}(r)-V_{cl}(R)$. Here $p=4.39073,\ j=0.328333,\ q=1.47436 $, corresponding to an horizon radius $R=1$ and to charge and angular momentum chemical potentials respectively $\mu= 0.997878$ and $\Omega=-1.96861$. On the left $p_{\psi_a} = 8.2,\ p_{\f_a} = 9.3$, while on the right $p_{\psi_a} = 10,\ p_{\f_a} = 9.3$. For $\Omega<-2\fr{p_{\psi_a}}{p_{\f_a}}$ the minimum is stable, as the energy is less than that at the horizon, signaling the onset of an instability.}
    \label{fig:GGT11instability}
    \end{figure}

    The effective potential is now obtained by a Legendre transform
    \begin{equation}\label{GGT11pot}
        \begin{split}
            V_{cl}(r)&=p_{\psi_a} \dot\psi_a+p_{\f_a}\dot\f_a-\cL_{eff},\\
            &= \fr 1{2\,b^2}\sqrt{W \l r^2 \l b^2+\l p_{\psi_a}+ A_{\s_3}\r^2\r+4\,b^2\l p_{\f_a}^2-p_{\psi_a}^2\r\r}-\l2+f\r\, p_{\psi_a}+\l A_t-f A_{\s_3}\r.
        \end{split}
    \end{equation}
    Notice that 
    \begin{equation}
        p_{\phi_a}\geq p_{\psi_a},
    \end{equation}
    and that for $p_{\phi_a}= p_{\psi_a}$ the potential simplifies to
    \begin{equation}\label{GGT11potequalcharges}
        V_{cl}^{p_{\f_a}=p_{\psi_a}}(r)= \fr r{2\,b^2}\sqrt{W \l b^2+\l p_{\psi_a}+ A_{\s_3}\r^2\r}-\l2+f\r\, p_{\psi_a}+\l A_t-f A_{\s_3}\r.
    \end{equation}
     We choose a gauge such that $A$ is regular at the horizon, namely we take $A_t\mapsto A_t-\mu$.\\ In this way we, since $\mu=A_t(R)-f(R) A_{\s_3}(R)$ and $W(R)=0$, the potential at the horizon reads
     \begin{equation}
         V_{cl}(R)=-\l2+\Omega\r p_{\psi_a},
     \end{equation}
     where $\Omega=f(R)$ is the chemical potential associated with angular momentum. The potential is characterized by two interesting behaviors, showed in Figure \ref{fig:GGT11instability}. We can either have a stable or an unstable minimum outside the horizon. The appearance of stable minima can be linked to the onset of angular momentum superradiance.
    
\subsubsection{The limit of large radius and angular momenta}
    We take $r\sim \sqrt{p_{\f_a}}\sim  \sqrt{p_{\psi_a}}\sim \ve  $ in the $\ve\to \infty$ regime. The potential reduces to
    \begin{equation}
    \begin{split}
        V_{cl}(r)=&2  ^2 (p_{\f_a}-p_{\psi_a})\ve+\left(\frac{r^2}{4 p_{\f_a}}+\frac{p_{\f_a}}{r^2}\right)-\fr1{\ve^2}\left(   \fr{16 j^2 p p_{\psi_a}^2+8 j p_{\psi_a} q r^2-r^4}{8p_{\f_a}r^4}+\right. \\
        &\left.- \fr{4 j p p_{\psi_a}-2 j p_{\psi_a} q+q r^2}{r^4}+\fr1{4r^4} p_{\f_a} (8 (p- q)+1)+\fr{r^4 }{64p_{\f_a}^3}     \right)+\cO\left(\fr1{\ve^3}\right),
    \end{split}
    \end{equation}
    which is minimized at
    \begin{equation}
        r_{min}=\sqrt{2\,p_{\f_a}}-\frac{(p_{\f_a}-j p_{\psi_a}) (2 p (p_{\f_a}-j p_{\psi_a})-3p_{\f_a} q)}{2 \sqrt{2}p_{\f_a}^{5/2} \ve ^2},
    \end{equation}
    where its value is
    \begin{equation}
        V_{cl}(r_{min})=2 (p_{\f_a}-p_{\psi_a})\ve^2+1-\frac{(p_{\f_a}-j p_{\psi_a}) (p (p_{\f_a}-j p_{\psi_a})-2 p_{\f_a} q)}{2 p_{\f_a}^3 \varepsilon ^2}+\cO\left(\frac{1}{\varepsilon^3 }\right).
    \end{equation}
    By comparison with the value of the potential at the horizon, we argue that
    \begin{itemize}
        \item for $\Omega<-2\fr{p_{\f_a}}{p_{\psi_a}}$, 
        $$V_{cl}(r_{min})<V_{cl}(R),$$
        the minimum is stable and the brane is emitted\\
        \item for $\Omega>-2\fr{p_{\f_a}}{p_{\psi_a}}$, 
        $$V_{cl}(r_{min})>V_{cl}(R),$$
        the minimum is unstable, thus there is no signal of black hole instabilities.\\
    \end{itemize}
    We can link this to the onset of angular momentum superradiance, as, in these coordinates, the chemical potential $\Omega$ is minus the sum of the equal horizon angular velocities in the two planes of the $S^3\subset \text{AdS}_5$, which we call $\Omega_{1,2}$. The first solutions to go unstable are the ones with $p_{\f_a}=p_{\psi_a}$, and $\Omega<-2$ corresponds to $\Omega_1=\Omega_2>1$, which is exactly the condition for the onset of angular momentum superradiance. This is consistent with the thermodynamic discussion in Section \ref{Intro}.

\subsubsection{The supersymmetric limit}
\begin{figure}[h!]
        \centering
        \includegraphics[width=0.5\linewidth]{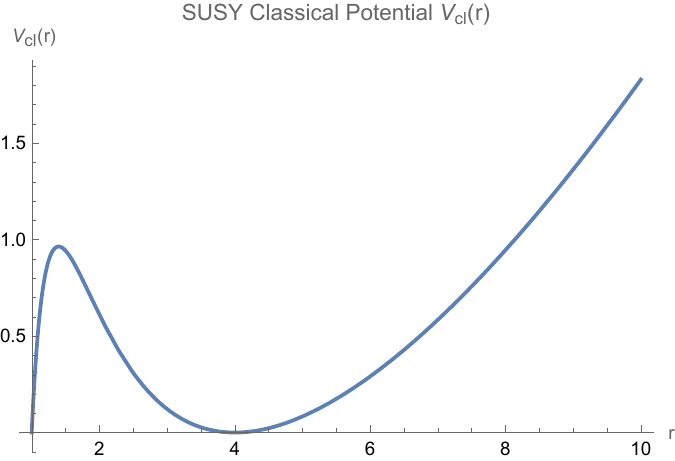}
        \caption{Effective potential experienced by the probe \D3-brane around a supersymmetric black hole for $R=1,\,\,p_{\psi_a}=p_{\f_a}=8.5$. We can see a BPS minimum appearing, thus signaling the presence of some marginally stable bound state.}
        \label{fig:GGT11SUSYclpot}
    \end{figure}

    We make the simplifying assumption\footnote{This is the relevant scenario since the $\kappa$-symmetry analysis forces the BPS minimum to be at $\dot\f_a=\dot\psi_a=0$. Notice that in inverting \eqref{GGT11momdef} to derive \eqref{dotintermsofp} we have made the implicit assumption that $p_{\psi_a} \neq p_{\phi_a}$. If we release this assumption we cannot use directly the second equation of \eqref{dotintermsofp}, but we need to start from scratch and manipulate \eqref{GGT11momdef}.} $p_{\f_a}=p_{\psi_a}$ and take the supersymmetric limit in the background solution. In this case, the potential reduces to \eqref{GGT11potequalcharges} and, using \eqref{5dGRsolsusycond}, it can be written as
    \begin{equation}\label{GGT11susypot}
    \begin{split}
        V_{cl}(r)=&\frac{r^2-R^2}{4r^6+4r^2R^6-R^8}\left( 4r^2\sqrt{\l 1 + r^2 + 2 R^2\r\l4 p_{\psi_a}^2 r^2 +  r^4 -2p_{\psi_a}R^4+ R^6\r }+\right.\\
        &+ 2 (2p_{\psi_a} + r^2) R^4 -4 (1 + 2p_{\psi_a}) r^4 - 8p_{\psi_a} r^2 R^2- 2 R^6\Bigg).
    \end{split}
    \end{equation}
    In Figure \ref{fig:GGT11SUSYclpot} we show the typical shape of the potential with BPS minimum.\\
    
    \nin
    We know from the $\kappa$-symmetry analysis of Appendix \ref{ksymmetry} that $\dot\psi_a=\dot\f_a=0$ at the supersymmetric minimum. By imposing this condition on \eqref{GGT11momdef}, one finds
    \begin{equation}\label{GGT11susymincond}
        p_{\psi_a}=\frac{r^2+R^2}{2},
    \end{equation}
    which is precisely the condition for the BPS minimum to appear. We can conclude that a marginally stable, BPS minimum appears outside the black hole horizon for
    \begin{equation}
        p_{\psi_a}>  R^2.
    \end{equation}

\subsubsection{Asymptotically flat limit}

    We want to study the asymptotically flat limit in the $p_{\f_a}=p_{\psi_a}$ sector. We restore the units by taking     \begin{equation}
        r\to\fr r L,\quad t\to \fr tL,\quad p\to\fr p{L^2},\quad q\to \fr q{L^2},\quad j\to \fr j L, \quad A\to \fr AL.
    \end{equation}
    If we consider the $L\to+\infty$ limit\footnote{The BMPV solution \cite{Breckenridge:1996is}, with usual dependence on the two parameters $\mu$ and $j$\footnote{Where $\mu$ must not be confused with the charge chemical potential. }, is recovered by taking $p\to0,\quad q\to-\mu$ and a standard choice of coordinates can be obtained by defining $\rho^2=r^2-\mu,\quad a=\fr{j\mu}2$.}, the five-dimensional metric and gauge field are \eqref{5dsolution} with 
    \begin{equation}
        W(r)  =1-\frac{1}{r^2}\left(2 p-2 q\right)+\frac{1}{r^4}\left(q^2+2pj^2\right).
    \end{equation}
    The potential\footnote{Reinstating units, \eqref{GGT11potequalcharges} reads $$V_{cl}(r)= \fr r{2\,L^2\,b^2}\sqrt{W \l b^2+\l p_{\psi_a}+ A_{\s_3}\r^2\r}-\l2+f\r\fr{p_{\psi_a}}L+A_t -\fr f L A_{\s_3}$$} \eqref{GGT11potequalcharges}  reduces to
    \begin{equation}
    \begin{split}
        V_{lim}(r)=& \fr{
 r\sqrt{(j q - 2 p_{\psi_a} r^2)^2 (2 j^2 p + q^2 - 2 p r^2 + 
     2 q r^2 + r^4)}}{r^6 - j^2 (q^2 - 2 p r^2)}+\fr {
 2 j p_{\psi_a} (q^2 - 2 p r^2 + q r^2)}{-r^6 + j^2 (q^2 - 2 p r^2)}+\\
     &+\fr q{r^2} + \fr{
 j^2 q (q^2 - 2 p r^2 + q r^2)}{r^8 + j^2 (-q^2 r^2 + 2 p r^4)} + \fr{
 j^2 - \mu^3}{j^2 - \mu^4},
    \end{split}
    \end{equation}
    which goes to zero at infinity for all the values of the parameters\footnote{In the BMPV limit, the potential reduces to $$ V_{BMPV}(r)=\fr{j^2 (2 p_{\psi_a} + r) (r^2 - \mu) - 
 j (-1 + \mu) \mu^4 + \mu^3 (-r^3 - 
    2 p_{\psi_a} r^2 \mu + (2 p_{\psi_a} + r) \mu^2)}{(r^3 - 
   j \mu) (j^2 - \mu^4)},$$ which is also flat at infinity. We verified that taking $p\to0$ and $q\to-\mu$ and only after $L\to \infty$ gives the same result.}. Indeed, the confining nature of AdS revealed essential for keeping the brane at finite radius. 
    This is consistent with the interpretation in terms of superradiance: although superradiance can also occur for asymptotically flat black holes, it is the confining nature of AdS that turns the superradiant amplification into an instability.

\section{Generalization to AdS$_5 \times  Y^{p,q}$}\label{WrappedYpq}
It is straightforward to generalize the probe brane constructions to asymptotically AdS$_5\times Y^{p,q}$ backgrounds. The $Y^{p,q}$ are a class of Sasaki-Einstein manifolds, with topology $S^2\times S^3$. In general, they have $SU(2)\times U(1)\times U(1)$ symmetry and can be written as a $U(1)$ fibration over manifolds of topology $S^2\times S^2$. We consider the metric\footnote{Set $L=1$.} 
\begin{equation}
\begin{split}
     ds_{Y^{p,q}}^2=& \frac{1-y}{6}\l d\theta^2+\sin^2\theta d\f^2\r+\frac{1}{v\, w} dy^2 + \frac{v\,w}{36} \l d\b +\cos\theta d\f\r^2+\\
     &+\fr19\l d\psi-\cos\theta d\f+y\l d\b+\cos\theta d\f\r\r^2,
\end{split}
\end{equation}
for 
\begin{equation}
    v= \frac{a-3y^2+2y^3}{a-y^2},\quad w=\frac{2\l a-y^2\r}{1-y}.
\end{equation}
The class of manifold is parametrized by two coprime integers $p$ and $q$, in terms of which
\begin{equation}
    a=\fr12-\frac{p^2-3q^2}{4p^3}\sqrt{4p^2-3q^3},
\end{equation}
and $y_1\leq y\leq y_2$, for
\begin{equation}
    \begin{split}
        -1<y_1&= \frac{1}{4p}\l 2p-3q-\sqrt{4p^2-3q^2} \r<0,\\
        1>y_2&= \frac{1}{4p}\l 2p+3q-\sqrt{4p^2-3q^2} \r>0,
    \end{split}
\end{equation}
roots of 
\begin{equation}
    a-3y^2+2y^3=0.
\end{equation}
For $q\neq0$, the coordinate ranges are $\theta\in[0,\pi],\,\, \f\in[0,2\pi),\,\,y\in[y_1,y_2],\,\,\beta\in[0,2\pi\ell),\,\,\psi\in[0,2\pi)$, where $\ell=\fr q{3q^2-2p^2+p\sqrt{4p^2-3q^2}}$. If $q=0$, $Y^{p,0}\simeq \fr{T^{1,1}}{\mathbb{Z}_p}$, and the metric reduces to $ds_{T^{1,1}}^2$ as presented in \eqref{D3pureAdsT11} with $\psi\in[0,\fr{4\pi}p)$.
\subsection{The ten-dimensional uplift}
    We consider the uplifting ansatz given in \cite{Buchel:2006gb} and uplift the five-dimensional solution \eqref{5dsolution}:
    \begin{equation}
        \begin{split}
            ds_{10}^2 =& ds_5^2+\frac{1-y}{6}\l d\theta^2+\sin^2\theta d\f^2\r+\frac{1}{v\, w} dy^2 + \frac{v\,w}{36} \l d\b +\cos\theta d\f\r^2+\\
            &+\fr19\l d\psi-\cos\theta d\f+y\l d\b+\cos\theta d\f\r+3A\r^2,\\
            ds_5^2 =& -\frac{r^2 W(r)}{4b(r)^2}dt^2+\frac{dr^2}{W(r)} + \frac{r^2}{4}(\sigma_1^2+\sigma_2^2)+b(r)^2(\sigma_3+f(r)\, dt)^2,\\
                A =& \frac{q}{r^2}\left(dt -\frac{j}{2}\sigma_3 \right).
        \end{split}
    \end{equation}
    The five-form is
    \begin{equation}
        \begin{split}
            F_{(5)}=&\l 1+\star_{10}\r G_{(5)}\\
            G_{(5)}=& -4 \text{vol}_5+\fr13 J\wedge\star_5\,F,
        \end{split}
    \end{equation}
    for 
    \begin{equation}
        J=\frac{1-y}{6}\sin\theta d\theta\wedge d\f+\fr16dy\wedge\l d\b+\cos\theta d\f\r
    \end{equation}
    K\"ahler form and $F=dA$.\\
    The four-form Ramond-Ramond potential is
    \begin{equation}
        C_{(4)} = C_V- \frac{2}{27 }\psi \,dg_5 \wedge dg_5 -\frac{1}{6}\star_5 F\wedge g_5+\frac{1}{18} A\wedge dg_5 \wedge\l g_5+3 A\r,
    \end{equation}
    where
    \begin{equation}
        g_5=d\psi-\cos\theta d\f+y\l d\b+\cos\theta d\f\r,
    \end{equation}
    and $C_V$ was defined in \eqref{Cv}.
\subsection{The brane wrapped around the internal directions}
    Along the lines of Sec. \ref{GGT11}, we consider the background metric in the shifted coordinates \eqref{5dsolutionshifted}.
    The probe embedding is
    \begin{equation}
        \begin{split}
        t&=\tau, \quad r=const,\quad \theta_a=\theta_a(\tau),\quad\phi_a=\f_a(\tau),\quad \psi_a= \psi_a(\tau)\\
            \theta&=2\,\arctan\left(c\,\zeta^{m}\right),\quad \phi=m\,\vf,\quad \psi=\gamma, \quad \b=const, \quad y=y_{1,2},
    \end{split}
    \end{equation}
    while the world-volume action is \eqref{GGT11action}.\\
    The pullback of the Ramond-Ramond four-form potential gives
    \begin{equation}
        P[C_{(4)}]=-\fr29 \l1-y_{1,2}\r\l A_t +A_{\sigma_3} \l 2+\cos\theta_a\dot\f_a+\dot\psi_a\r\r \frac{c^2 m^2 \zeta^{2m-1}}{\left(1+c^2\zeta^{2m}\right)^2}d\tau\wedge d\zeta\wedge d\vf\wedge d\gamma,
    \end{equation}
    while the square root of the induced metric determinant is
    \begin{equation}
        \sqrt{-\det\l P[g_{\mu\nu}]\r}=\frac{c\,m^2 \zeta^{m-1}\sqrt{A\l 1+c^4\zeta^{4m}\r +B c^2\zeta^{2m}}}{9\sqrt{3}\left(1+c^2\zeta^{2m}\right)^2}\sqrt{\bar\Delta},
    \end{equation}
    where
    \begin{equation}
        A=a-3y^2+2y^3|_{y=y_{1,2}},\quad B= 12-2\,a -24 \,y+18y^2 -4y^3|_{y=y_{1,2}},
    \end{equation}
    and 
    \begin{equation}
        \begin{split}
            \bar\Delta=&\fr{r^2 W}{4 b^2}
-b^2 (2 + f)^2-\fr{r^2}{4}\dot{\theta}_a^{\,2}
-\fr18 \l
4 b^2+r^2+(4 b^2-r^2)\cos \l2\theta_a\r
\r\dot{\phi}_a^{\,2}+\\
&
-2 b^2 (2+f)\dot{\psi}_a-b^2\dot{\psi}_a^{\,2}
-2 b^2\cos\theta_a\,\dot{\phi}_a
\l2+f+\dot{\psi}_a\r.
    \end{split}
    \end{equation}
    Using $a-3y^2+2y^3|_{y=y_{1,2}}=0$, we get $A=0$ and $B=12(1-y_{1,2})^2$ and reduce to
    \begin{equation}
        \sqrt{-\det\l P[g_{\mu\nu}]\r}=\fr29(1-y_{1,2})\frac{c^2\,m^2 \zeta^{2m-1}}{\left(1+c^2\zeta^{2m}\right)^2}\sqrt{\bar\Delta}.
    \end{equation}
    By using
    \begin{equation}
       \int_0^\infty d\zeta\, \frac{c^2\,m^2 \zeta^{2m-1}}{\left(1+c^2\zeta^{2m}\right)^2}=\frac{|m|}{2},
    \end{equation}
    we obtain the effective Lagrangian
    \begin{equation}
        {\cal L}_{eff}=-\fr89\pi^2|m|T_3\l1-y_{1,2}\r\l \sqrt{\bar\Delta}+ A_t +A_{\sigma_3} \l 2+\dot\psi_a\r\r.
    \end{equation}
    Observe that, for $q=0$, $y_{1,2}=0$ and the result reduces to \eqref{GGT11lagrangian} for $(m_1,m_2)=(m,0)$ and $(c_1,c_2)=(c,0)$ since indeed $Y^{1,0}=T^{1,1}$. 

\subsection{The dual Giant Graviton}
    In this case, we consider the background \eqref{5dsolution} and the embedding
    \begin{equation}
        \begin{split}
        t=\tau, \quad r=const,\quad \theta_a=\z,\quad\phi_a=\vf,\quad \psi_a= \g,\quad
            \theta,\phi,\b,y=const,\quad \psi=\psi(\tau),
    \end{split}
    \end{equation}
    and evaluate the world-volume action \eqref{GGT11action}. For simplicity, we only consider the case of a single rotation cycle in the internal space, which correspond, in the main text analysis, to $q_n=q_c$.
    
    The pullback of the Ramond-Ramond four-form potential gives
    \begin{equation}
        P[C_{(4)}]=\fr{r^4-R^4}{8}d\tau\wedge d\zeta\wedge d\vf\wedge d\gamma,
    \end{equation}
    while the square root of the induced metric determinant is
    \begin{equation}
        \sqrt{-\det\l P[g_{\mu\nu}]\r}=\fr{r^2}{24}\sin\z\sqrt{\fr{r^2W}{b^2}9\l A_{\s_3}^2+b^2\r-36b^2\l A_t-A_{\s_3}f\r^2-4b^2\dot\psi\l 6\l a_t-A_{\s_3}f\r+\dot\psi\r}.
    \end{equation}
    The effective Lagrangian is\footnote{We set $T_3=\fr1{2\pi^2}$.}
    \begin{equation}
        {\cal L}_{eff}=-\fr{r^2}3\sqrt{\fr{r^2W}{b^2}9\l A_{\s_3}^2+b^2\r-36b^2\l A_t-A_{\s_3}f\r^2-4b^2\dot\psi\l 6\l a_t-A_{\s_3}f\r+\dot\psi\r} +r^4-R^4,
    \end{equation}
    which reproduces \eqref{DDBHT11lagrangian} for $\dot\theta_{1,2}=\dot\f_{1,2}=0$.

\section{Comparison to Giant Gravitons around AdS$_5 \times S^5$ black holes}\label{GGS5}

    In the previous section we considered \D3-branes wrapped on non-trivial cycles of the internal manifold $T^{1,1}$, where the topology of the Sasaki-Einstein space allows for supersymmetric configurations exhibiting BPS minima in the effective potential. In contrast, the five-sphere $S^5$ only contains contractible cycles. Giant Gravitons, which are \D3-branes wrapped on an $S^3\subset S^5$ and carrying angular momentum along a maximal circle of the internal space, can therefore minimize their energy by shrinking to zero size. We find that, at the onset of angular momentum superradiance, the black hole emits a Giant Graviton that reduces its energy by shrinking into a simple graviton, respecting the original Grey Galaxy construction of \cite{Kim:2023sig,Choi:2025lck}. On the other hand, in the supersymmetric limit, a line of degenerate minima connects the shrunken graviton to  marginally stable BPS minima at finite radius\footnote{Let us mention that related work in this direction was reported in \cite{Mondal:2025slz}, where the supersymmetry of various (dual) Giant Gravitons was studied in detail.}

    In the dual four-dimensional $\mathcal{N}=4$ super Yang-Mills theory, whose field content includes six real scalar fields $X^I$ transforming in the vector representation of the $SO(6)$ R-symmetry, Giant Gravitons correspond to determinant or subdeterminant operators constructed from a complex combination of these scalars. Introducing $Z = X^1 + iX^4$, a maximal Giant Graviton is dual to the operator $\det Z$, while smaller giants correspond to subdeterminant operators
    \begin{equation}
    \det_k Z = \frac{1}{k!}\epsilon_{i_1\ldots i_k a_{k+1}\ldots a_N}
    \epsilon^{j_1\ldots j_k a_{k+1}\ldots a_N}
    Z^{i_1}_{\,\,\,j_1}\cdots Z^{i_k}_{\,\,\,j_k},
    \end{equation}
    with $k\leq N$. The angular momentum of the brane along the circle in $S^5$ maps, through AdS/CFT, to the corresponding $U(1)$ R-charge of the operator.

    In the configurations that we consider, the brane also carries angular momentum along the Hopf fiber of the $S^3\subset \text{AdS}_5$. From the boundary perspective this corresponds to inserting the operator on the three-sphere where the field theory is defined and giving it orbital angular momentum under the spatial rotation group $SO(4)\simeq SU(2)_L\times SU(2)_R$. We will study these Giant Graviton configurations in the background of AdS$_5\times S^5$ black holes.

\subsubsection{Intermezzo: pure AdS$_5\times S^5$}\label{GGS5pure}
    It is physically insightful to consider the Giant Graviton in pure AdS$\times S^5$, where we are forced to place it in the center of AdS$_5$ and to make it move in the internal directions for it not to shrink. We report here the computations of \cite{Grisaru:2000zn}. We take the  metric
    \begin{equation}
        \begin{split}
            ds^2&=ds_{\text{AdS}_5}^2+ds^2_{S^5},\\
            ds_{\text{AdS}_5}^2&=-\l1+r^2\r dt^2+\frac{dr^2}{1+r^2}+r^2\l d\a_1^2+\sin^2(\a_1)\l d\a_2^2+\sin^2(\a_2) d\a_3\r\r,\\
            ds_{S^5}^2&=d\rho_s^2+\cos^2(\rho_s) d\f^2+\sin^2(\rho_s)\l d\c_1^2+\sin^2(\c_1)\l d\c_2^2+\sin^2(\c_2) d\c_3^2\r \r,
        \end{split}
    \end{equation}
    with coordinates $t\in \mathbb{R},\,\,r\in[0,\infty),\,\,\a_{1,2}\in[0,\pi],\,\,\a_3\in[0,2\pi),\,\,\rho_s\in[0,\fr\pi2],\,\,\f\in[0,2\pi),\,\,\chi_{1,2}\in[0,\pi],\,\,\chi_3\in[0,2\pi)$.
    The embedding is
    \begin{equation}
    \begin{split}
         t=\s^0,\quad r=const,\quad \a_{1,2,3}=const, \quad
        \rho_s=const,\quad \f=\f(\s^0),\quad \c_i=\s^i\quad i=1,2,3.
    \end{split}
    \end{equation}
    The four-form potential is 
    \begin{equation}
    \begin{split}
         C_{(4)}=&-r^4\sin^2(\a_1)\sin(\a_2) dt\wedge d\a_1\wedge d\a_2\wedge d\a_3+\\
         &+\sin^4(\rho_s)\sin^2(\a_1)\sin(\a_2) d\f\wedge d\c_1\wedge d\c_2\wedge d\c_3.
    \end{split}
    \end{equation}
    We extract the Lagrangian from the action 
    \begin{equation}
         \begin{split} 
            S= \frac{N}{2 \pi^2 }\left(  -\int d^4\s \sqrt{-{\rm det}(P[g_{\mu\nu}])}+   \int P[C_{(4)}] \right), \\
            \end{split}
            \end{equation} 
    and obtain
    \begin{equation}
        \cL_{eff}=-N\sin^3(\rho_s)\l \sqrt{1+r^2-\cos^2(\rho_s)\dot\f^2}-\sin(\rho_s)\dot\f  \r,
    \end{equation}
    where $\sin\l\rho_s\r$ is the Giant Graviton size.
    By taking $\dot\f=0$ and changing sign, we reduce to the effective potential
    \begin{equation}
        V_{cl}(r,\rho_s)=N\sin^3(\rho_s)\sqrt{1+r^2}
    \end{equation}
    We can now clearly see the topological difference between $S^5$ and $T^{1,1}$. In pure AdS$_5\times T^{1,1}$, a D3-brane wrapped on a non-contractible cycle (see Section \ref{GGT11}) is stable at the center of AdS. By contrast, in pure AdS$_5\times S^5$ a D3-brane wrapping an $S^3\subset S^5$ can minimize its energy by shrinking to zero size, as signaled by the overall $\sin^3\rho_s$ factor in the potential.
    
    In order to have a finite size, stable Giant Graviton, we have to allow for motion in the internal directions. We define the momentum
     \begin{equation}\label{GGpureS5momdef}
         N p_\f= \frac{\de\cL_{eff}}{\de\dot\f}=N\sin^3(\rho_s)\l\sin(\rho_s)+\frac{\cos^2(\rho_s)\dot\f}{\sqrt{1+r^2-\cos^2(\rho_s)\dot\f^2}}\r,
     \end{equation}
     and solve for $\dot\f$
     \begin{equation}
         \dot\f=\frac{p_\f-\sin^4(\rho_s)}{\cos(\rho_s)}\sqrt{\frac{1+r^2}{p_\f^2-2p_\f\sin^4(\rho_s)+\sin^6(\rho_s)}}.
     \end{equation}
     The effective potential we reduce to is
     \begin{equation}
         V_{cl}(r,\rho_s)=\fr N{\cos(\rho_s)}\sqrt{\frac{1+r^2}{p_\f^2-2p_\f\sin^4(\rho_s)+\sin^6(\rho_s)}}\l p_\f^2+\fr{1-4p_\f-\cos(2\rho_s)}{2} \sin^4(\rho_s)\r.
     \end{equation}
    As in the static scenario, the Giant Graviton is forced to the AdS$_5$ center $r=0$, where the potential simplifies to
    \begin{equation}
        V_{cl}(\rho_s)=N\sqrt{p_\f^2+\tan^2(\rho_s)\l p_\f-\sin^2(\rho_s)\r^2}.
    \end{equation}
    \begin{figure}[h!]
        \centering
        \includegraphics[width=0.5\linewidth]{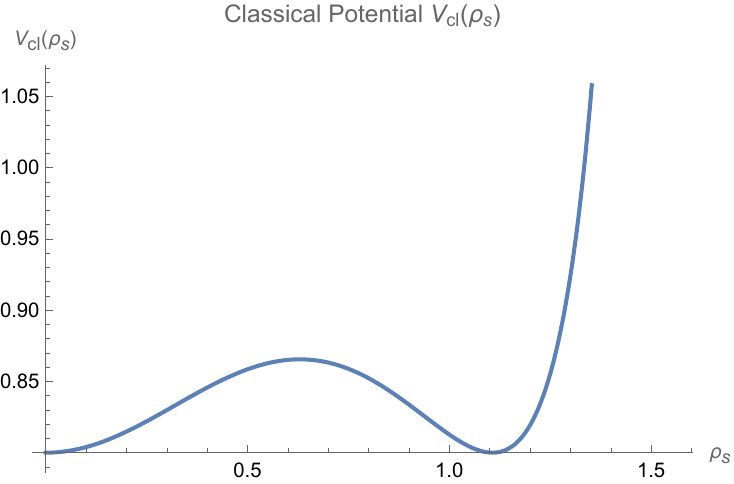}
        \caption{Effective potential for a Giant Graviton orbiting in a maximal cycle of $S^5\subset \text{AdS}_5\times S^5$ with momentum $p_\f=0.8$ . We set $ N=1$ and $r=0$.}
        \label{fig:GGpureS5rhopot}
    \end{figure}
    
    \nin
    For non-vanishing momentum, a degenerate minimum appears at finite Giant Graviton size $\rho_s>0$. We report the potential in Figure \ref{fig:GGpureS5rhopot}.
    By setting $\dot\f=1$ in \eqref{GGpureS5momdef}, we obtain the condition for the minimum to appear
    \begin{equation}
        p_\f=\sin^2(\rho_s).
    \end{equation}
    We can thus see that even in the pure AdS$_5\times S^5$ background, a Giant Graviton with finite size can exist at the center of AdS$_5$ if rotating with the speed of light in the internal space.

\subsection{The effective Lagrangian for black hole backgrounds}
    Once again, we consider the background metric in the shifted coordinates \eqref{5dsolutionshifted} and the embedding
    \begin{equation}
    \begin{split}
        t=&\tau, \quad r=const,\quad \psi_a=\psi_a(\tau),\quad \phi_a=\phi_a(\tau),\quad \theta_a=\theta_a(\tau),\\
        \rho_s =& const, \qquad \theta_s=\s^1,\quad \f_s=\s^2,\quad \z_s,\psi_s=\s^3,
    \end{split}
    \end{equation}
    which, in the supersymmetric case and for null angular velocities, reduces to the one presented in Appendix E of \cite{Aharony:2021zkr}. As in the conifold truncation, by considering directly the embedding with null angular velocities, one obtains a flat potential. Instead, in order to arrive to the correct Routhian formulation of the problem, we generalize the embedding of \cite{Aharony:2021zkr} to include angular momenta. We are now in presence of a Giant Graviton, a D3-brane wrapped around an $S^3\subset S^5$, moving in the internal directions and rotating in the black hole background \eqref{GGS510dmetric}.
    In order to extract the effective Lagrangian from the world-volume action 
    \begin{equation}\label{GGS5action}
            \begin{split} 
            S= \frac{N}{2 \pi^2 }\left(  -\int d^4\s \sqrt{-{\rm det}(P[g_{\mu\nu}])}+   \int P[C_{(4)}] \right), \\
            \end{split}
            \end{equation} 
    we compute the pull-back of the four-form potential
    \begin{eqnarray}
        P[C_{(4)}]=-\fr18 \sin\l\s_1\r\sin^4(\rho_s)\l A_t+\l 2+\dot \psi_a+\cos\theta_a \dot \f_a\r A_{\s_3}\r d\s^0\wedge d\s^1\wedge d\s^2\wedge d\s^3,
    \end{eqnarray}
    and the square root of minus the determinant of the pull-back of the metric
    \begin{equation}
        \sqrt{-{\rm det}(P[g_{\mu\nu}])}=\fr{\sin(\s^1)\sin^3(\rho_s)}8 \sqrt{ \Delta},
    \end{equation}
    where
    \begin{equation}
        \begin{split}
            \Delta=& 1/8 \left(-4 \left((2 A_{\s_3} + A_t)^2 + 2 b^2 (2 + f)^2\right) + \fr{2 r^2 W}{b^2} - 
       2 r^2 \dot\theta_a^2 +\right.\\
             &\left.- 
       4 \cos\l2\rho_s\r (2 A_{\s_3} + A_t + 
          A_{\s_3} \dot\psi_a)^2 - 
       4 \dot\psi_a \left(2 A_{\s_3} (2 A_{\s_3} + A_t) + 
          4 b^2 (2 + f) + (A_{\s_3}^2 + 2 b^2) \dot\psi_a\right) +\right.\\
          &\left.- 
       8 \cos\theta_a \dot\f_a (A_{\s_3} \left(2 A_{\s_3} + A_t) + 2 b^2 (2 + f) + 
          A_{\s_3} (2 A_{\s_3} + A_t) \cos\l2\rho_s\r + (A_{\s_3}^2 + 2 b^2 + A_{\s_3}^2 \cos\l2\rho_s\r) \dot\psi_a\right)+\right.\\
          &\left.- \left(2 A_{\s_3}^2 + 4 b^2 + r^2 + 
          4 A_{\s_3}^2 \cos\l2\rho_s\r \cos\theta_a^2 + (2 A_{\s_3}^2 + 4 b^2 - 
             r^2) \cos\l2\theta_a\r\right) \dot\f_a^2 +\right)\\
        \end{split}
    \end{equation}
    After integrating we obtain 
    \begin{equation}
        \begin{split}
            \cL_{eff}= &-N\sin^3(\rho_s)\l \sqrt{ \Delta}+\sin(\rho_s) \l  A_t+\l 2+\dot \psi_a+\cos\theta_a \dot \f_a\r A_{\s_3}\r\r.
        \end{split}
    \end{equation}
    By setting the angular velocities to zero and imposing supersymmetry, one obtains a vanishing Lagrangian.\\
    Since the black hole has an $SU(2)$ symmetry, we can use it again to simplify the motion. Following Section \ref{Motion_on_the_squahed_T11}, we set
    \begin{equation}
        p_{\theta_a}=0,\quad \theta_a=\arccos\l\fr{p_{\psi_a}}{p_{\f_a}}\r,
    \end{equation}
    and define the conserved momenta
    \begin{equation}\label{GGS5momdef}
    \begin{split}
         N\ppa &=\frac{\de \cL_{eff}}{\de\dot\psi_a},\quad  N \pfa=\frac{\de \cL_{eff}}{\de\dot\f_a}.
    \end{split}
    \end{equation}
    In order to solve for $\dot\psi_a$ and $\dot \f_a$, we take 
        \begin{equation}
            \fr{\text{II}\fr{p_{\f_a}}{p_{\psi_a}} - \text{I}}{\text{I}+ A_{\s_3} \sin^4\l\rho_s\r},
        \end{equation}
        where I and II are respectively the first and second equations of \eqref{GGS5momdef}, solve for $\dot\psi_a$ and reinsert in II. The equations and the solutions are articulated and not particularly illuminating. We avoid writing them down.
\subsection{The effective potential}
\begin{figure}[h!]
            \centering
            
            \vspace{-2cm}

            \hspace{.1cm}
            
            \begin{subfigure}[b]{\linewidth}
                \centering
                \includegraphics[width=0.8\linewidth]{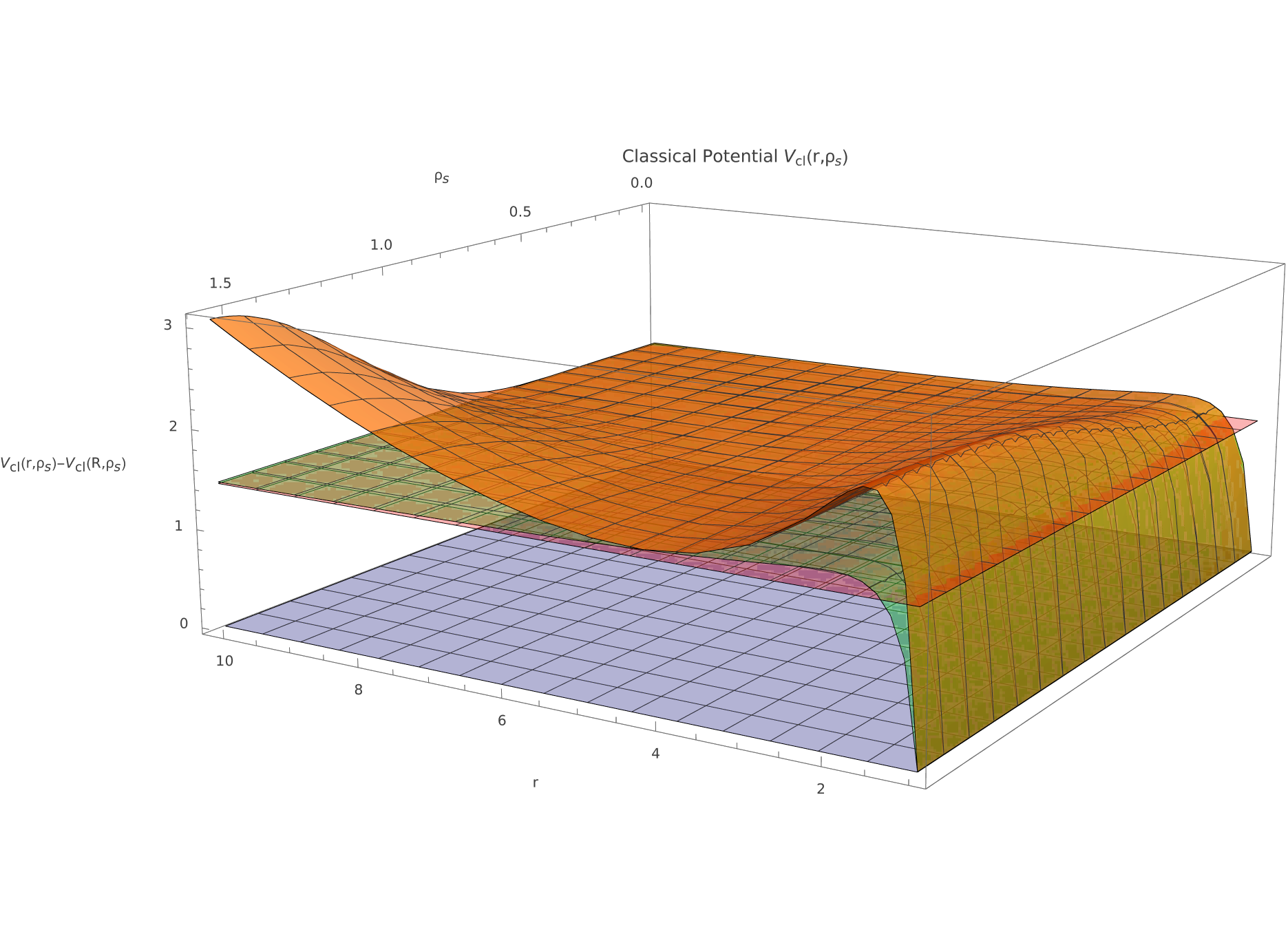}
                
            \end{subfigure}
            
            \vspace{-2cm}

            \hspace{-1cm}
            
            \begin{subfigure}[b]{
            \linewidth}
                \centering
                \includegraphics[width=0.8\linewidth]{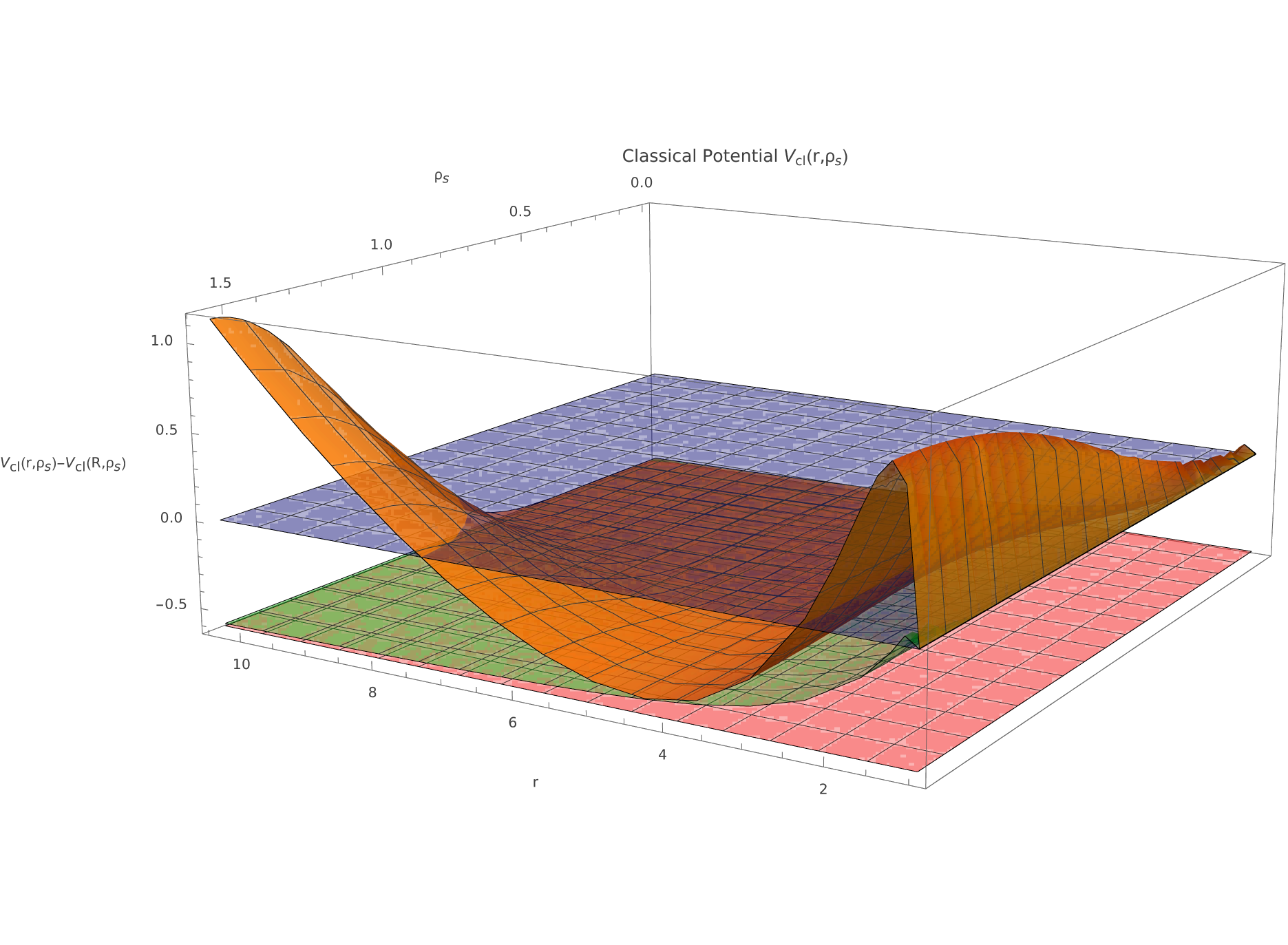}
                
            \end{subfigure}
            \caption{In orange the energy of the probe at distance $r$ minus that at the horizon $V_{cl}(r,\rho_s)-V_{cl}(R,\rho_s)$, in blue the zero, in green the potential at zero Giant Graviton radius $V_{cl}(r,0)-V_{cl}(R,0)$, while in pink we plot the energy of the probe at zero radius of the Giant Graviton and at large $r\sim1.2\cdot 10^3$. Here $p=4.46628,\ j=0.353333,\ q=1.44165 $, corresponding to an horizon radius $R=1$ and to charge and angular momentum chemical potentials respectively $\mu= 0.916699$ and $\Omega=-2.06113$. On top $p_{\psi_a} = 10,\ p_{\f_a} = 9$, while on the bottom $p_{\psi_a} = 10,\ p_{\f_a} = 10$. For $\Omega<-2\fr{p_{\psi_a}}{p_{\f_a}}$ the black hole gains energy by emitting the dual giant, but, as we can see from the fact that the pink surface is lower than the orange minimum, the Giant Graviton gains energy by shrinking to zero size and moving to large radius.  }
            \label{fig:GGS5pot}
        \end{figure}

    The effective potential is obtained by Legendre transform
    \begin{equation}
    \begin{split}
        V_{cl}=&N \l\dot\psi_a\ppa+\dot\f_a\pfa\r-\cL_{eff}\\
        =& \fr{\fr{\sqrt{\tilde\Delta}}{b r} - 
     8 \ppa (2 b^2 (2 + f) + 2 A_{\s_3} (2 A_{\s_3} + A_t) \cos\l\rho_s\r^2) + 
     16 b^2 (A_t - A_{\s_3} f) \sin\l\rho_s\r^4}{16 (b^2 + A_{\s_3}^2 \cos\l\rho_s\r^2)},
        \end{split}
    \end{equation}
    for 
    \begin{equation}
        \begin{split}
            \tilde\Delta=& \left(4 b^4 \left(A_t - A_{\s_3} f\right)^2 - \left(A_{\s_3}^2 + 2 b^2\right) r^2 W + \left(4 b^4 \left(A_t - A_{\s_3} f\right)^2 - 
          A_{\s_3}^2 r^2 W\right) \cos\l2\rho_s\r\right) \\
          &\left(-64 \left(A_{\s_3}^2 + 2 b^2\right) \left(\pfa - \ppa\right) \left(\pfa + 
          \ppa\right) - 
       2 \left(5 \left(A_{\s_3}^2 + b^2\right) + 12 A_{\s_3} \ppa + 
          16 \ppa^2\right) r^2 +\right.\\
          &\left.+\left(-64 A_{\s_3}^2 \left(\pfa - \ppa\right) \left(\pfa + 
             \ppa\right) + \left(15 \left(A_{\s_3}^2 + b^2\right) + 32 A_{\s_3} \ppa\right) r^2\right) \cos\l2\rho_s\r+ \right.\\
             &\left.+   r^2 \left(-2 \left(3 \left(A_{\s_3}^2 + b^2\right) + 4 A_{\s_3} \ppa\right)\cos\l4\rho_s\r + \left(A_{\s_3}^2 + b^2\right)\cos\l6\rho_s\r\right)\right).
        \end{split}
    \end{equation}
    Note that for $\rho_s=\fr\pi2$, i.e. for maximal size of the Giant Graviton, the potential simplifies to \eqref{GGT11pot}. The typical shapes of the potential, when admitting minima, are reported in Figure \ref{fig:GGS5pot}. We take $N=1$. \\
    For $A_t\mapsto A_t-\mu$, the potential at the horizon reads
    \begin{equation}
        V_{cl}(R,\rho_s)=-\ppa \l2+\Omega\r.
    \end{equation}  
    
\subsubsection{The limit of large radius and angular momenta}
     Consider the limit $\ve\to\infty$ for $\ppa\sim\pfa\sim\sqrt{r}\sim \ve$. The potential reads
     \begin{equation}
         V_{cl}(r,\rho_s)=2\l\pfa-\ppa\r\ve^2+ \fr{\pfa}{r^2} \l 1-A^2\cos^2\l\rho_s\r\r + A\sin^4\l\rho_s\r+\fr{r^2}{4 \pfa} \sin^6\l\rho_s\r+\cO\l\fr1{\ve^2}\r,
     \end{equation}
     where
     \begin{equation}
         A=q\fr{j^2- \l1-j^2\r R^2}{j^2q+R^4}.
     \end{equation}
     Since at large $r$, the $\cO(1)$ term has always positive derivative with respect to $\rho_s$, the energy is always decreased by going at large $r$ and $\rho_s=0$, as one can see from Figure \ref{fig:GGS5pot}. The brane, after being emitted at the onset of angular momentum superradiance, shrinks to zero size into a simple graviton.

\subsubsection{The supersymmetric limit}
\begin{figure}[h!]
    \centering
    \includegraphics[width=0.8\linewidth]{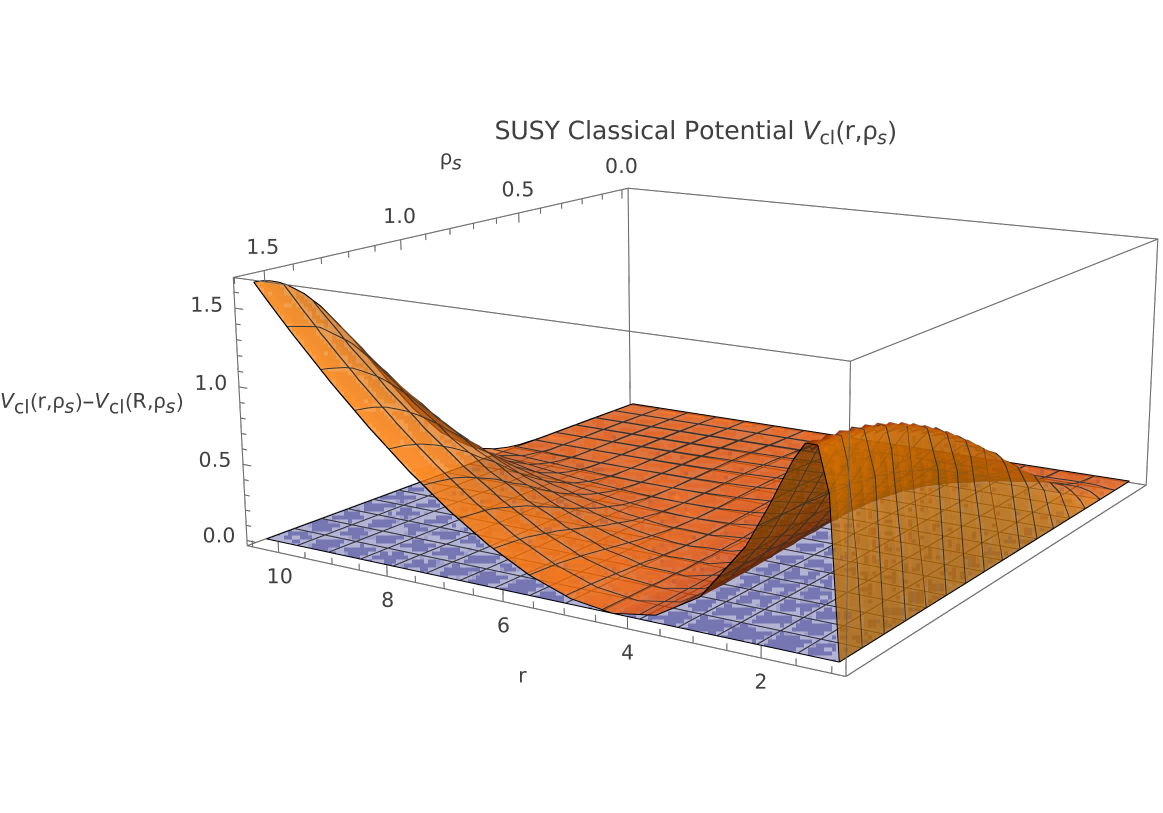}
    \caption{In orange the energy of the probe at distance $r$ minus that at the horizon $V_{cl}(r,\rho_s)-V_{cl}(R,\rho_s)$ in the supersymmetric limit, in blue the zero. We take $R=1$  and  $p_{\psi_a} = p_{\f_a} = 10$. The charge and angular momentum chemical potentials are respectively $\mu= 1$ and $\Omega=-2$.  We see a line connecting the zero at  null Giant Graviton size to degenerate marginally stable BPS minima at finite Giant Graviton size.}
    \label{fig:GGS5susymin2}
\end{figure}

    In the supersymmetric limit, for $\ppa=\pfa$, the potential reduces to
    \begin{equation}
        \begin{split}
            V_{cl}(r,\rho_s)= &N\frac{r^2-R^2}{4r^6+4r^2R^6-R^8\sin^2(\rho_s)}\Biggl(4\ppa R^4 \sin^2(\rho_s) -2\l2r^4-r^2 R^4+R^6\r \sin^4(\rho_s)+\\
            &-8 r^2 \ppa \l r^2+R^2\r+4 \,r\sqrt{  r^4 + 2 r^2 R^2+ r^2\sin^2(\rho_s)+R^4\cos^2(\rho_s) } {\,\,\,}\cdot\\
            &\cdot \sqrt{4\ppa^2r^2-2\ppa R^4\sin^4(\rho_s)+\l r^4+R^6 \r \sin^6(\rho_s)} \Biggr).
        \end{split}
    \end{equation}
    Observe that, in this case, the potential flattens for $\rho_s=0$ and a line of degenerate minima of null energy appears in the $(r,\rho_s)$ plane. We report the potential shape in Figure \ref{fig:GGS5susymin2}. 

    According to the $\kappa$-symmetry analysis performed in \cite{Aharony:2021zkr}, the probe is BPS for $\dot\psi_a=0$ and, indeed, by setting $\dot\psi_a=0$ in the momentum definition \eqref{GGS5momdef} we get the condition for a BPS minimum to appear
    \begin{equation}
        p_\F=\frac{r^2+R^2}{2}\sin^2(\rho_s).
    \end{equation}
    We thus can see that a Giant Graviton moving at the speed of light along a maximum cycle in the  $S^5$\footnote{The metric \eqref{GGS510dmetric} has a component $\fr13\l d\psi_a+A_t dt\r^2$, which, being $A_t\to-1$ at $r\to \infty$, gives angular speed $-1$ to the Giant Graviton along the $S^5$ fiber, which is a maximal cycle.} and with momentum along the $S^3\subset \text{AdS}_5$ fiber can have a stable BPS configuration at finite radius $\rho_s>0$.

\section{Conclusions and implications for the dual CFT}

    In this work, we studied superradiant instabilities of AdS$_5$ black holes via a probe D3-brane analysis. Inspired by the construction of \cite{DDBH}, we have found stable minima at the onset of charge superradiance in AdS$_5\times T^{1,1}$ and, to a certain extent, in AdS$_5\times Y^{p,q}$ backgrounds. 
    One concern remains regarding the charge condensation instability, as charged AdS black holes are expected to develop scalar hair. While numerical hairy black hole solutions are known in the AdS$_5\times S^5$ truncation and were shown to be unstable to charge superradiance \cite{DDBH}, no corresponding solutions have so far been constructed in the conifold truncation. It would be desirable to construct these solutions and check whether they are unstable as well.

    We also applied the same technique to the study of angular momentum superradiance. According to the conjecture of \cite{Kim:2023sig}, the endpoint of this instability is a Grey Galaxy, namely a black hole surrounded by a gas of gravitons. Our D3 probe brane approach provides insight into the microscopic composition of the gas. For the $S^5$ case the Giant Graviton is allowed to shrink, and it reduces to an ordinary graviton, reproducing the picture proposed in \cite{Kim:2023sig,Choi:2025lck}. Indeed the effective potential for the Giant Graviton develops a continuous family of degenerate, marginally stable BPS minima, connecting between a flat direction at vanishing brane size and a minimum at maximal size. For the $T^{1,1}$ case,
    a D3-brane wrapping internal directions is topologically stabilized by non-contractible cycles and supersymmetric bound states composed of the black holes and baryonic branes exist at the onset of superradiance.

    Our BPS bound states for $T^{1,1}$ are consistent with supersymmetric realizations of both the Dual Dressed Black Holes and the Grey Galaxies, at least within the probe approximation. Provided supersymmetry is not broken beyond the probe approximation, these BPS composite systems should play a role in the supersymmetric microstate counting in the dual CFT. As mentioned previously, the existence of these bound states was already predicted by \cite{Choi:2026faq}\footnote{See also \cite{Chang:2024zqi,Chen:2025sum,Chang:2025wgo,Behan:2025hbx,Belin:2025hsg,Kim:2026rnx,Giusto:2026rpl} for recent developments of the fortuity program in supersymmetric gauge theories with holographic dual description.}, from the dual field theory side, which distinguished two possible black hole dressings from the analysis of fortuitous cohomologies in the conifold. In one case one has a dual Giant Graviton wrapping a cycle in AdS$_5$, corresponding to the field theory image of the dual dressed black hole of \cite{DDBH, Choi:2025lck}, across which the RR five-form flux jumps by one unit, with negligible backreaction at leading order in $N$. In the other case one has a baryonic condensate of D3-branes wrapping a non-contractible internal cycle, where the flux is unchanged between UV and IR and the backreaction is significant already at leading order in $N$.

 In their setup, bound states with baryonic condensate interpolate between an  asymptotic AdS$_5\times T^{1,1}$ region in the UV and the AdS$_5\times S^5$ throat in the IR. The condensate arises from a macroscopic baryonic Higgs VEV: at large VEV the two regions in the IR are weakly coupled, like the probes studied in this paper, while for small VEV a baryonic hairy BPS black hole in AdS$_5\times T^{1,1}$ is expected. The latter, which consists of the fully backreacted supergravity solution where the two can no longer be treated separately (i.e. at generic VEV) is still to be constructed. In addition, \cite{Choi:2026faq}  observed that different dressings of the fortuitous operators have distinct physical interpretations.  For example dressings by baryonic letters are associated with excitations of the D3-brane condensate itself. On the other hand dressings by mesonic letters correspond to  BPS excitations around the  black hole, in strict analogy with the Grey Galaxy picture. It would be interesting to sharpen the dictionary between these cohomological dressings and the probe bound states constructed here, especially focusing on  the probe regime of a small number of D3-branes as a dilute limit of the  backreacted, macroscopic baryonic condensate involving all $N$ colors.

 Let us mention that one motivation for considering composite systems of probe D-branes was to explore the possible presence of multi-center AdS$_5$ configurations in an extreme mass ratio limit. A related interpretation on the dual field-theory side might involve the insertion of a defect, along the lines of \cite{Gaiotto:2012xa}. Establishing the corresponding fully backreacted solutions would help clarify the relation between these two descriptions.

Let us finally mention a few other extensions of the framework studied here. Recently there was some interest in D3-brane configurations around the Gutowski-Reall black hole and their possible role as non-perturbative contributions to the superconformal index for theories on AdS$_5 \times SE_5$; see, for example, \cite{Mondal:2025slz,Deddo:2025lfm,BenettiGenolini:2026hmz}. Most of these works focus on D3-branes wrapping Euclidean three-cycles, following the approach of \cite{Aharony:2021zkr}, where such configurations are interpreted as non-perturbative saddle points contributing to the index. In particular, \cite{Aharony:2021zkr} also studied the Giant Graviton embedding in the AdS$_5 \times S^5$ black hole background and showed that the configuration preserves supersymmetry\footnote{A different probe D3-brane,  extended across the time and
a radial direction, and wrapped on one compact direction in AdS$_5$ and one compact
direction in $S^5$, was considered in \cite{Chen:2023lzq} (see also \cite{Cabo-Bizet:2023ejm,Amariti:2024bsr}) . Such brane was interpreted as a half
BPS Gukov-Witten surface on the dual side.}.

Our configurations may appear as additional saddle points contributing to the superconformal index already at $\sim\cO(N^2)$ in the large-$N$ limit. Indeed in \cite{Choi:2025lck,Deddo:2025jrg} numerical evaluations of the superconformal index at $N\simeq 10$ revealed deviations from the contribution expected from the single-center supersymmetric black hole. These deviations were found to be qualitatively consistent with the emergence of additional phases, including supersymmetric Grey Galaxies and Dual Dressed Black Holes, in $\mathcal{N}=4$ Super Yang--Mills theory. The search for similar phenomena in Klebanov-Witten theory, where the structure of the superconformal index remains comparatively less understood \cite{Benini:2020gjh}, would be very interesting in this regard.

    Another intriguing perspective on these bound states comes from recent studies of complex saddles in the Euclidean gravitational path integral \cite{Krishna:2026rma,BenettiGenolini:2025jwe,BenettiGenolini:2026raa}. In particular, the analysis of \cite{Krishna:2026rma} suggests that phases such as Grey Galaxies and Dual Dressed Black Holes may emerge precisely at the onset of divergences in the path integral. Clarifying this connection would require the construction of Euclidean counterparts of these configurations and a detailed investigation of their role as gravitational saddles.

    We hope to return to some of these questions in future work.

\begin{comment}

\end{comment}

\section*{Acknowledgements}

The authors would like to thank Riccardo Argurio, Connor Behan, Iosif Bena, Francesco Benini, Nikolay Bobev, Frederik Denef, Ohad Mamroud for useful discussions and correspondence that lead to this work. AA and RS acknowledge funding from Italian Ministero dell’Istruzione, Universit\`a e Ricerca (MIUR), in part by Istituto Nazionale di Fisica Nucleare (INFN) through the “Gauge Theories, Strings, Supergravity” (GSS) research project. CT acknowledges support from the Belgian Fonds National de la Recherche Scientifique (FNRS) via the MISU grant 40024018 ”Pushing horizons in Black Hole physics” and the Ministerio de Ciencia, Innovación y Universidades / Agencia Estatal de Investigación (AEI), via the Ramon y Cajal program, RYC2024-048886-I. RS was supported by the “Thesis Abroad” program of Università di Milano and thanks the Department of Theoretical and Mathematical Physics at Université Libre de Bruxelles for its hospitality.

\begin{appendices}
\addtocontents{toc}{\protect\setcounter{tocdepth}{1}}

\section{$\kappa$-symmetry analysis}\label{ksymmetry}
\subsection{Killing spinor}\label{KS}
    We follow Appendix C of \cite{Aharony:2021zkr} and find the Killing spinor for the black hole \eqref{5dsolution} uplifted to AdS$_5\times T^{1,1}$.
\subsubsection{Type IIB supergravity }
     Consider the five-dimensional minimal gauged supergravity black hole solution, with associated Killing spinor \cite{Aharony:2021zkr}, uplifted to type IIB supergravity on AdS$_5\times SE_5$, with $SE_5$ a generic Sasaki-Einsten manifold. We write the bosonic action for the metric $G_{MN}$ and the five-form $F_{(5)}$ 
     \begin{equation}
         S_{IIB}\sim\int d^{10}x \sqrt{-G}\, \l \cR_{10} -\frac{1}{4\cdot5!} F_{M_1...M_5}F^{M_1...M_5}\r,
     \end{equation}
     where $F_{(5)}=\star_{10} \,F_{(5)}$ is the self-duality condition. We consider the Sasaki-Einstein manifold $SE_5$ as a $U(1)$ fibration over a base $B$, and write the ansatz \cite{Aharony:2021zkr}
     \begin{equation}\label{KSgeneralreductionansatz}
         \begin{split}
             ds_{10}^2=ds_5^2+ds_B^2+\l e^9\r ^2,&\quad e^9=\fr 13 \l d\psi+\cA +3 A\r,\\
             F_{(5)}=\l 1+\star_{10}\r \,G_{(5)},&\quad G_{(5)}=-4\e_{(5)}+ J\wedge \star_5\, F,
        \end{split}
     \end{equation}
    where $J$ is the K\"hler form such that 
    \begin{equation}
        \fr12J\wedge J = \text{vol}_B,
    \end{equation}
    and $\cA$ is a $U(1)$ connection on $B$ such that
    \begin{equation}
        d\cA=6 J.
    \end{equation}
    Here $ds_5^2$ is the five-dimensional (black hole) spacetime metric with volume form $\e_{(5)}$ and Hodge dual operator $\star_5$.
    We have
    \begin{equation}
        \star_{10} \, G_{(5)}= 2 J\wedge J\wedge e^9-F\wedge J\wedge e^9.
    \end{equation}
    A possible choice for the four-form potential $C_{(4)}$ such that $F_{(5)}=dC_{(4)}$ is 
    \begin{equation}
        C_{(4)}=-4\b_{(4)} +\fr13\tilde\cA\wedge \l \l J-\fr12 F\r\wedge e^9+\fr12\star_5 F\r,
    \end{equation}
    where we defined $\b_{(4)}$ via $\e_{(5)}=d\b_{(4)}$ and $\tilde\cA$ via $d\tilde\cA=6J$.
    The restriction to the metric and five-form flux automatically imposes the vanishing of the ten-dimensional dilatino (the fermionic supersymmetric partner of the scalar dilaton) variation, while giving for the gravitino
    \begin{equation}\label{KSgravitvar}
        \d_\ve \psi_M =\nabla_M\ve +\frac{i}{16\cdot5!}F_{P_1...P_5}\Gamma^{P_1...P_5}\,\Gamma_M\ve,
    \end{equation}
    where $\ve$ is the ten-dimensional spinor. Consider the following decomposition for the gamma matrices
    \begin{equation}
        \Gamma^M=\{ \g^\mu \otimes\mathbb{I} \otimes\s_1,\,\mathbb{I}\otimes\h \g^{\ua}\otimes \s_2\},
    \end{equation}
    where $\mu=0,...,4$ are reserved for the spacetime metric, $\ua=5,...,9$ are the directions of the  Sasaki-Einstein internal manifold, with $\ua=5,...,8$ directions of the base $B$, and $\s_{1,2}$ are Pauli matrices.
    Take \cite{Aharony:2021zkr}
    \begin{equation}
        \g^{01234}=-i,\quad \h\g^{56789}=1, \quad \G_{11} =\G^0\cdots\G^9=\mathbb{I}\otimes\mathbb I\otimes \s_3,
    \end{equation}
    and decompose the spinor in 
    \begin{equation}\label{KSdecomposition}
        \ve=\e \,\otimes\,\c \,\otimes\begin{pmatrix}
            0\\1
        \end{pmatrix}.
    \end{equation}
    The latter satisfies
    \begin{equation}
        \G_{11}\ve=-\ve,
    \end{equation}
    i.e $\s_3\ve=-\ve$.

\subsubsection{The Killing spinor for $T^{1,1}$}
    Given the covariant derivative $\h\nabla$ on $SE_5$, the internal manifold admits killing spinor $\c$ satisfying \cite{Aharony:2021zkr}
    \begin{equation}\label{KSequnSE}
        \l\h\nabla_{\ua}+\fr i2\h\g_{\ua}\r\c=0,
    \end{equation}
     and the projectors
    \begin{equation}
        \h\g^{56}\c=\h\g^{78}\c=i\c,\quad \h\g^9\c=-\c.
    \end{equation}
    While in \cite{Aharony:2021zkr} the authors specify to the case of $SE_5=S^5$, we will consider $SE_5=T^{1,1}$ (the argument holds for general $Y^{p,q}$ manifolds). See the $T^{1,1}$ as a $U(1)$ fibration over $S^2\times S^2$. The coordinates are $\theta_{1,2}=[0,\pi]$, $\f_{1,2}=[0,2\pi)$, $\psi=[0,4\pi)$, and the vielbein can be written
    \begin{equation}
        \begin{split}
        &e^5=\fr1{\sqrt{6}}d\theta_1,\quad e^6=\fr1{\sqrt{6}}\sin\theta_1d\f_1,\quad \cA=\cos\theta_1d\f_1+\cos\theta_2d\f_2,\\
        &e^7=\fr1{\sqrt{6}}d\theta_2,\quad e^8=\fr1{\sqrt{6}}\sin\theta_2d\f_2,\quad e^9=\fr13\l d\psi+\cA+3A\r.
        \end{split}
    \end{equation}
    The K\"ahler form is
    \begin{equation}
        J=-\fr16\l\sin\theta_1 d\theta_1\wedge d\f_1+\sin\theta_2 d\theta_2\wedge d\f_2\r.
    \end{equation}
    Equation \eqref{KSequnSE} reduces to
    \begin{equation}
        \begin{split}
            \l\sqrt 6 \cot\theta_{1,2}\l\de_\psi -\fr i2\r -\sqrt6 \csc\theta_{1,2}\,\de_{\f_{1,2}}  \r \c&=0,\\
            \de_{\theta_{1,2}}\c&=0,\\
            \l\de_\psi-\fr i2\r \c&=0,
        \end{split}
    \end{equation}
    which are solved by the spinor
    \begin{equation}\label{KST11}
        \c=e^{\fr i2 \psi}\c_0,
    \end{equation} 
    with $\c_0$ constant spinor such that
    \begin{equation}
        i\h\g^{56}\c_0=i\h\g^{78}\c_0=\h\g^9\c_0=-\c_0.
    \end{equation}

\subsubsection{The Killing spinor for the five-dimensional black hole}
    By using the results presented above in \eqref{KSgravitvar} one obtains the gravitino variation of five-dimensional minimal gauged supergravity
    \begin{equation}\label{KSMGSvariation}
        \l \nabla_\mu-\fr32iA_\mu -\fr12 \g_\mu -\frac{i}{8}\l\g_\mu^{\nu\rho}-4\d_\mu^\nu\g^\rho\r F_{\nu\rho}\r\e=0.
    \end{equation}
   Consider the Gutowsli-Reall black hole solution we introduced in \eqref{5dGRsolution}
   \begin{equation}
        \begin{split}
            ds_5^2  &=  - \tf^2 dt^2 - 2 \tf^2\, \Psi\, dt \sigma_3 + U^{-1} dr^2 + \frac{r^2}{4} \left( \sigma_{1}^2 + \sigma_{2}^2 +\Sigma\,  \sigma_3^2 \right),\\
            A &=  \left(  \frac{R^2}{r^2} -1\right)dt- \frac{R^4}{4\, r^2}\sigma_3 ,
        \end{split}
    \end{equation}
    where
    \begin{equation}
        \begin{split}
          U &= \left( 1 - \frac{R^2}{r^2} \right)^2 \left( 1 + 2 R^2 + r^2\right), \quad
        \Sigma = 1 +\frac{R^6}{r^4} - \frac{R^8}{4 \, r^6},\\
        \tf &= 1 - \frac{R^2}{r^2}, \quad
        \Psi = -\frac{r^2}{2} \left( 1 + \frac{2 R^2}{r^2} + \frac{3 R^4}{2 r^2 (r^2- R^2)}\right),
        \end{split}
    \end{equation}
    and the $\s_i$ are defined in \eqref{sigmas}.\\    
    In order to obtain the metric in orthotoric coordinates of \cite{Aharony:2021zkr}, we perform the coordinate change 
    \begin{equation}
        r\mapsto 2\, \sqrt{R^2+\tilde \xi},\quad \theta_a\mapsto\arccos \eta,\quad \psi_a\mapsto\tilde \F,\quad \f_a\mapsto \tilde\Psi,
    \end{equation}
    and get
    \begin{equation}\label{BH metric orthotoric a=b}
        \begin{split}
            ds^2&= - \tilde f_a^2 \bigl( dt - \omega \bigr)^2 + \frac1{\tilde f_a} \left( \frac{\tilde\xi}{\tilde\cF} \, d\tilde\xi^2 + \frac{\tilde\cF}{\tilde\xi} \bigl( d\tilde\Phi + \eta\, d\tilde\Psi \bigr)^2 + \tilde\xi \left( \frac{d\eta^2}{1-\eta^2} + (1-\eta^2) \, d\tilde\Psi^2 \right) \right),\\
            A &= -\tilde f_a\, dt -\frac{\tilde\xi}{\tilde f_a}\l1-\tilde f_a\r^2 \bigl( d\tilde\Phi + \eta\, d\tilde\Psi\bigr) ,
        \end{split}
    \end{equation}
    where \footnote{We report here the different conventions with respect to \cite{Aharony:2021zkr} for better comparison. They consider more general non extremal supersymmetric black holes, so, in their equations (C.35)-(C.42) we set $\tilde m=0$. We also have a different definition of the gauge one-form $A_{here}=\fr23 A_{there}$. We thus set $\a=\frac{3}{2}$ and factor out a global $\fr32$ in (C.42). We set $a=j=\frac{R^2}{2}\l1+\frac{R^2}{2}\r^{-1}$. Here $\tilde f_a$ is their $f$, while $\tilde f$ is defined in \eqref{5dGRfuncdef}. We also list some relations between the two conventions:
    \begin{equation}
    \begin{split}
        &\tilde\xi=\frac{r^2}{4}\l1-\frac{R^2}{r^2}\r,\quad  \tilde f_a= \frac{2\l1-a\r\tilde \xi }{a+2\l1-a\r\tilde \xi}=1-\frac{R^2}{r^2},\\ &A_{\s_3}=-\frac{a^2}{4\l1-a\r^2}\frac{\tilde f_a}{\tilde\xi}=-\frac{\tilde\xi}{\tilde f_a}\l1-\tilde f_a\r ^2=-\frac{R^4}{4 r^2},\\
        &\tilde\cF = \tilde\xi^2\l 1+4\tilde\xi +6 \frac{a}{1-a}\r=\frac{r^4}{16}\left( 1 - \frac{R^2}{r^2} \right)^2 \left( 1 + 2 R^2 + r^2\right),\\
        &\omega=\l \frac{1}{3\tilde f_a}\l \frac{\tilde\cF'(\tilde\xi)}{2\tilde\xi}-1\r -\frac{a^2}{4(1-a)^2\tilde\xi}\r \l d\tilde\F+\eta \,d\tilde\Psi\r = \frac{r^2}{2} \l 1+\frac{2 R^2}{r^2}+\frac{3R^4}{2r^2\l r^2-R^2\r}\r\s_3.
    \end{split}
    \end{equation}}
    \begin{equation}
        \begin{split}
            \tilde f_a =\tilde f,\quad \tilde \cF=\frac{r^4}{16}\,U,\quad \omega=-\Psi\,( d\tilde\Phi + \eta\, d\tilde\Psi\bigr).
        \end{split}
    \end{equation}
     Observed that here we choose a gauge where the electrical potential vanishes at the horizon. When considering the Dual Dressed Black Holes, instead, we prefer a gauge where the gauge field vanishes at infinity, where 
     \begin{equation}
         A_t=-\tilde f_a+1.
     \end{equation}
        Choose the vielbein
 \begin{equation}\label{vielbein orthotoric a=b}
     \begin{split}
         e^0 &= \tilde f_a (dt - \omega) ,\quad
e^1 = -\frac1{\tilde f_a^{1/2}} \sqrt{ \frac{\tilde\xi}{\tilde\cF(\tilde\xi)}} \, d\tilde\xi ,\quad
e^2 = \frac1{\tilde f_a^{1/2}} \sqrt{ \frac{\tilde\cF(\tilde\xi)}{\tilde\xi}} \, \bigl( d\tilde\Phi + \eta\, d\tilde\Psi \bigr), \\
e^3 &= - \frac1{\tilde f_a^{1/2}} \sqrt{ \frac{\tilde\xi}{1-\eta^2} }\, d\eta ,\quad
e^4 = - \frac1{\tilde f_a^{1/2}} \sqrt{\l1-\eta^2\r \, \tilde\xi }\, d\tilde\Psi .
     \end{split}
 \end{equation}
The spinor solving \eqref{KSMGSvariation} is\footnote{We have $\a=\fr32$.}
\begin{equation}\label{KSspacetime}
    \e=e^{\fr i2   \tilde\F}\sqrt{\tilde f_a}\, \e_0,
\end{equation}
with constant $\e_0$ satisfying
\begin{equation}
    i\e=-\g^0\e=-\g^{12}\e=\g^{34}\e.
\end{equation}

%\end{comment}
\subsubsection{Ten-dimensional Killing spinor}

We can now use \eqref{KST11}, \eqref{KSspacetime} and \eqref{KSdecomposition} to construct the ten-dimensional killing spinor, which takes the form
\begin{equation}
    \ve=e^{\fr i 2 \l\tilde\F +\psi\r}\sqrt{\tilde f_a}\,\,\ve_0,
\end{equation}
with $\ve_0$ satisfying
\begin{equation}\label{KSprojections}
    \G^{09}\ve=\ve,\quad \G^{12} \ve=-i\ve,\quad \G^{34}\ve=\G^{56}\ve=\G^{78}\ve =i\ve.
\end{equation}
Observe that in the coordinates of \eqref{5dGRsolution} the spinor reads
\begin{equation}\label{KSspacetimeGR}
    \ve=e^{\fr i 2 \l\psi_a +\psi\r}\sqrt{\tilde f_a}\,\,\ve_0,
\end{equation}
while in the coordinates of \eqref{5dGRsolMinwallacoords} it reads
\begin{equation}\label{KSspacetimeGRMinwalla}
     \ve=e^{\fr i 2 \l\psi_a-2t +\psi\r}\sqrt{\tilde f_a}\,\,\ve_0,
\end{equation}

\subsection{Probe D3-brane embeddings}\label{KSymm}
    Supersymmetry for probe D3-brane embeddings in absence of world-volume fluxes is enforced by imposing the $\kappa$-symmetry condition \cite{Aharony:2021zkr} 
    \begin{equation}
        \Theta\ve=\mp i\ve,
    \end{equation}
    where 
    \begin{equation}\label{KSTheta}
        \Theta=\frac{1}{24} \frac{\e^{\a_1...\a_4}}{\sqrt{-h}} \frac{\de X^{\mu_1}}{\de\s^{\a_1}}\cdots\frac{\de X^{\mu_4}}{\de\s^{\a_4}} e^{M_1}_{\,\,\,\,\,\,\mu_1}\cdots e^{M_4}_{\,\,\,\,\,\,\mu_4}\, \G_{M_1...M_4},
    \end{equation}
    and $\mp$ corresponds to branes or anti-branes. 
    We defined $\s^{0,...,3}$ and $X^\mu(\s)$ the world-volume coordinates and embedding of the brane, $\a=0,...,3$ and $\mu=\{t,...,\psi\}$ respectively the world-volume and ten-dimensional spacetime indices, while $M=0,...,9$ are flat ten-dimensional indices. We follow \cite{Aharony:2021zkr} and take $\e^{0123}=1$. The induced metric on the brane is 
    \begin{equation}\label{KSindmetric}
        h_{\a_1\a_2}=\frac{\de X^{\mu_1}}{\de\s^{\a_1}}\frac{\de X^{\mu_2}}{\de\s^{\a_2}}g_{\mu_1\mu_2},
    \end{equation}
    $h$ is the determinant.

\subsubsection{Wrapped branes}

    Consider the embedding \eqref{GGT11embedding} describing a D3-brane wrapping the two independent two-cycles of $T^{1,1}$ respectively $m_1$ and $m_2$ times. In the conventions adopted in this section, the embedding reads
    \begin{equation}\label{KSgiantgravitonembedding}
    \begin{split}
         t= \s^0,\quad \tilde\xi,\eta,\tilde \F,\tilde\Psi= \mathrm{const},\quad
         \theta_{1,2}=2\arctan \l c_{1,2 } \l\s^{1\,}\r^{m_{1,2}} \r  ,\quad \f_{1,2}=m_{1,2}\, \s^2,\quad \psi=\s^3.
    \end{split}
    \end{equation}
    Using \eqref{KSindmetric} and \eqref{KSTheta}, we find
    \begin{equation}
        \sqrt{-h}=\fr29\l \frac{c_1^2 m_1^2 \l\s^1\r^{2m_1-1}}{\left(1+c_1^2\l\s^1\r^{2m_1}\right)^2}+ \frac{c_2^2 m_2^2 \l\s^1\r^{2m_2-1}}{\left(1+c_2^2\l\s^1\r^{2m_2}\right)^2}\r \tilde f_a,
    \end{equation}
    and
    \begin{equation}
    \begin{split}
        \Theta=&-\frac{2}{9\sqrt{-h}} \left( \frac{c_1^2 m_1^2 \l\s^1\r^{2m_1-1}}{\left(1+c_1^2\l\s^1\r^{2m_1}\right)^2}\tilde f_a \,\G^{0569}+ \frac{c_2^2 m_2^2 \l\s^1\r^{2m_2-1}}{\left(1+c_2^2\l\s^1\r^{2m_2}\right)^2}\tilde f_a \,\G^{0789}+\right.\\
        &\,\left. + \frac{c_1 m_1 c_2 m_2 \l\s^1\r^{m_1+m_2-1}}{\left(1+c_1^2\l\s^1\r^{2m_1}\right)\left(1+c_2^2\l\s^1\r^{2m_2}\right)} \tilde f_a\l\G^{0589}-\G^{0679}\r \right).
    \end{split}
    \end{equation}
    Making use of the projections \eqref{KSprojections}, one finds
    \begin{equation}
        \begin{split}
            &\G^{0589}\ve=\G^{0958}\ve=i\G^{0957}\ve=\G^{0967}\ve=\G^{0679}\ve,\\
            &\G^{0569}\ve=\G^{0956}\ve=i\G^{09}\ve=i\ve,\\
            &\G^{0789}\ve=\G^{0978}\ve=i\G^{09}\ve=i\ve,
        \end{split}
    \end{equation}
    so that all terms proportional to $\G^{0589}-\G^{0679}$ vanish when acting on the Killing spinor. Consequently,
    \begin{equation}
        \Theta\ve=-i\ve.
    \end{equation}
    The $\kappa$-symmetry condition is therefore satisfied, and the probe brane embedding \eqref{GGT11embedding} preserves supersymmetry.\\
    \begin{comment}

    \nin
    In contrast, the coordinate system used in \eqref{5dGRsolMinwallacoords} would require an embedding of the form
    \begin{equation}\label{KSgiantgravitonembedding}
    \begin{split}
         t&= \s^0,\quad \tilde\F=\f_0-2\s^0,\quad \tilde\xi,\eta,\tilde\Psi= const,\\
         \theta_{1,2}&=2\arctan \l c_{1,2 } \l\s^{1\,}\r^{m_{1,2}} \r  ,\quad \f_{1,2}=m_{1,2}\, \s^2,\quad \psi=\s^3,
         \end{split}
    \end{equation}
    for some constant $\f_0$. In this case, by making use of the projectors \eqref{KSprojections}, we can reduce to 
    \begin{equation}
        \Theta=-\frac{2\sqrt{\tilde\xi\,\tilde F}\,\,\G^{2789}+\tilde\xi\tilde f_a\sqrt{\tilde f_a}\l1-2\Psi\r\G^{0789}}{\sqrt{\tilde\xi\l\tilde\xi\tilde f_a^3\l1-2\Psi\r^2-4\tilde F\r}},
    \end{equation}
    which cannot satisfy the $\kappa$-symmetry condition.This embedding therefore fails to preserve supersymmetry\footnote{ This is why, when wrapping the internal directions, we take $\psi_a\mapsto\psi_a+2 t$ in the main text.}.
    \end{comment}

\subsubsection{Wrapped branes in AdS$_5\times Y^{p,q}$}
    Analogous results hold for $Y^{p,q}$ manifolds, see \cite{Buchel:2006gb}. In this case, the probe embedding is
     \begin{equation}
         \begin{split}
         t= \s^0,\quad \tilde\xi,\eta,\tilde \Phi,\tilde\Psi= \mathrm{const},\quad
         \theta=2\arctan \l c \l\s^{1\,}\r^{m} \r ,\quad \f=m\, \s^2,\quad \psi=\s^3,\quad y=y_{1,2},
    \end{split}
    \end{equation}
    while the projector reduces to
    \begin{equation}
        \Theta=-\frac{2}{9\sqrt{-h}} \l 1-y_{1,2} \r \frac{c^2 m^2 \l\s^1\r^{2m-1}}{\left(1+c^2\l\s^1\r^{2m}\right)^2}\tilde f_a \,\Gamma^{0569},
    \end{equation}
    where
    \begin{equation}
        \sqrt{-h}=\frac{2}{9} \l 1-y_{1,2} \r \frac{c^2 m^2 \l\s^1\r^{2m-1}}{\left(1+c^2\l\s^1\r^{2m}\right)^2} \tilde f_a.
    \end{equation}
    By using the projections \eqref{KSprojections}, one can find
    \begin{equation}
        \Theta \varepsilon = -i\varepsilon.
    \end{equation}
    This result reduces to the above $T^{1,1}$ case for $(p,q)=(1,0)$, $(c_1,c_2)=(c,0)$, $(m_1,m_2)=(m,0)$.
\subsubsection{dual Giant Gravitons}
    We now turn our attention to the dual Giant Graviton embedding, a \D3-brane wrapping the $S^3\subset \text{AdS}_5$. We will follow \cite{Aharony:2021zkr} and consider it static in the $T^{1,1}$ directions. The embedding is specified by
    \begin{equation}
        t=\s^0,\quad \eta=\s^1,\quad \tilde\F=\s^2-2\s^0, \quad \tilde \Psi=\s^3,\quad \psi,\theta_{1,2},\f_{1,2}=\mathrm{const}
    \end{equation}
    which leads to
    \begin{equation}\label{KST11DGGconst}
        \sqrt{-h}=-\tilde\xi\l A_{\s_3}+\tilde f_a \Psi\r, \quad \Theta =-\G^{0349}+\sqrt{\frac{\tilde \xi\tilde F}{-h\,\tilde f_a}}\l\G^{0234}-\G^{2349} \r.
    \end{equation}
    Using the projections \eqref{KSprojections} we find
    \begin{equation}\label{DGGprojections}
    \begin{split}
         &\G^{0234}\ve=-\G^{2340}\ve=\G^{2349}\ve,\\
         &\G^{0349}\ve=\G^{0934}\ve=i\ve,
    \end{split}
    \end{equation}
    and therefore
    \begin{equation}
        \Theta\ve=-i\ve.
    \end{equation}
   This confirms that the embedding preserves supersymmetry.    \\

    \nin
    In order to verify the supersymmetry of \eqref{DDBHT11embedding}, we need to consider the more general embedding
    \begin{equation}
        \begin{split}
            t&=\s^0,\quad \eta=\s^1,\quad \tilde\F=\s^2-2\s^0, \quad \tilde \Psi=\s^3,\\
            \psi&=\psi(\s^0),\quad\theta_{1,2}=\theta_{1,2}(\s^0),\quad\f_{1,2}=\f_{1,2}(\s^0),
        \end{split}
    \end{equation}
    which in general does not respect the $\kappa$-symmetry condition. Note that in this case we take 
    \begin{equation}
        A_t=-\tilde f_a+1.
    \end{equation}
 In Section \ref{spacetime_filling} we have seen that we need to impose
    \begin{equation}\label{qn=qc}
        q_n^2=6\l p_{\f_1}^2+p_{\f_2}^2 \r-3p_{\psi}^2=q_c^2=9p_\psi^2
    \end{equation}
     and we can preserve the $\mathbb {Z}_2$ symmetry of the internal manifold by imposing 
     \begin{equation}
         p_{\f_1}=p_{\f_2},\quad \text{i.e.} \quad \dot\f_1=\dot\f_2.
     \end{equation}
     Together with \eqref{qn=qc}, this implies
     \begin{equation}\label{equalvel}
         p_{\f_1}=p_{\f_2}=p_\psi ,\quad \text{i.e.} \quad \dot\f_1=\dot\f_2=\dot\psi.
     \end{equation}
    By considering \eqref{DDBHT11momentadef} and  \eqref{DDBHT11pmomentasol} with the supersymmetric constraints \eqref{5dGRsolsusycond}, we see that for\footnote{Note that in \cite{DDBH} is enforced the analogous condition $\dot\f_1=\dot\f_2=\dot\f_3=1$, where the change of sign is due to the definition of the $\f_i$'s.}
    \begin{equation}
        \dot\f_1=\dot\f_2=\dot\psi=-1
    \end{equation}
    the momenta definitions \eqref{DDBHT11momentadef} reduce to
    \begin{equation}
        -3 p_\psi=q_c=\frac{2 r^4-r^2R^4+R^6}{2\l r^2+R^2\r},
    \end{equation}
    which is the condition for the supersymmetric minima to occur. \\
    In conclusion, at the supersymmetric minima, one can consider the embedding
    \begin{equation}
        \begin{split}
            &t=\s^0,\quad \eta=\s^1,\quad \tilde\F=\s^2-2\s^0, \quad \tilde \Psi=\s^3,\\
            &\psi=-\s^0,\quad\theta_{1,2}=0,\quad\f_{1,2}=-\s^0,
        \end{split}
    \end{equation}
    and the $\kappa$-symmetry condition reduces to 
    \begin{equation}
        \Theta\ve =\l-\G^{0349}+\fr{r\,\sqrt{1+2R^2+r^2}}{r^2+R^2}\l\G^{0234}-\G^{2349} \r\r\ve=-i\ve,
    \end{equation}
    where we used \eqref{DGGprojections} to prove supersymmetry.

    \subsubsection{dual Giant Gravitons in AdS$_5\times Y^{p,q}$}
    We take the embedding
    \begin{equation}
        t=\s^0,\quad \eta=\s^1,\quad \tilde\F=\s^2-2\s^0, \quad \tilde \Psi=\s^3,\quad \theta,\f,\b,y=\mathrm{const}, \quad\psi=\psi(\tau),
    \end{equation}
    which, for\footnote{One could also consider the more general embedding, where $\b=c_1 \tau$, $\f=c_2\tau$, $\psi=c_3\tau$. In order for the $\kappa$-symmetry condition to hold, $c_3=c_2(1-y)-yc_1-3$. In the conifold, by setting $\theta_{1,2}=0$ and allowing for motion in the angular directions, which is equivalent to setting $q_n=q_c$, one obtains a Lagrangian dependent on $\dot\psi+\dot\f_1+\dot\f_2$. At the BPS minimum, they are all $-1$, meaning that the effective sum is $-3$. Imposing $\dot\psi=-3$ is analogous in this case to defining a "collective" coordinate.  } $\dot\psi=-3$, leads to
    \begin{equation}\label{KST11DGGconst}
        \sqrt{-h}=-\tilde\xi\l A_{\s_3}+\tilde f_a \Psi\r, \quad \Theta =-\G^{0349}+\sqrt{\frac{\tilde \xi\tilde F}{-h\,\tilde f_a}}\l\G^{0234}-\G^{2349} \r.
    \end{equation}
    Using the projections \eqref{KSprojections} we find
    \begin{equation}\label{DGGprojections}
    \begin{split}
         &\G^{0234}\ve=-\G^{2340}\ve=\G^{2349}\ve,\\
         &\G^{0349}\ve=\G^{0934}\ve=i\ve,
    \end{split}
    \end{equation}
    and therefore
    \begin{equation}
        \Theta\ve=-i\ve.
    \end{equation}
   This confirms that the embedding preserves supersymmetry.

  \end{appendices}
 \bibliographystyle{utphys}
  \bibliography{references}

\end{document}